\documentclass[twocolumn]{revtex4}

\usepackage{ragged2e}
\usepackage{amsmath}
\usepackage{graphicx}
\usepackage{rotating}

\usepackage[usenames,dvipsnames]{color}

\definecolor{darkblue}{RGB}{0,0,196}

\begin{document}

\title{Thermal freeze-out parameters and pseudo-entropy from charged hadron
spectra in high energy collisions \vspace{0.5cm}}

\author{Xu-Hong~Zhang$^{1,}$\footnote{xhzhang618@163.com; zhang-xuhong@qq.com},
Ya-Qin~Gao$^{2,}$\footnote{gyq610@163.com; gaoyaqin@tyust.edu.cn},
Fu-Hu~Liu$^{1,}$\footnote{Correspondence: fuhuliu@163.com;
fuhuliu@sxu.edu.cn},
Khusniddin~K.~Olimov$^{3,}$\footnote{Correspondence:
khkolimov@gmail.com; kh.olimov@uzsci.net}}

\affiliation{$^1$Institute of Theoretical Physics, State Key
Laboratory of Quantum Optics and Quantum Optics Devices \&
Collaborative Innovation Center of Extreme Optics, Shanxi
University, Taiyuan 030006, China
\\
$^2$Department of Physics, Taiyuan University of Science and
Technology, Taiyuan 030024, China
\\
$^3$Laboratory of High Energy Physics, Physical-Technical
Institute of Uzbekistan Academy of Sciences, Chingiz Aytmatov str.
$2^b$, 100084 Tashkent, Uzbekistan}

\begin{abstract}

\vspace{0.5cm}

\noindent {\bf Abstract:} We collected the transverse momentum
(mass) spectra of charged hadrons ($\pi^{-}$, $\pi^{+}$, $K^{-}$,
$K^{+}$, $\overline{p}$, and $p$) produced in collisions over a
center-of-mass energy range from 2.70 to 200 GeV (per nucleon
pair). The modified Tsallis--Pareto-type function (the TP-like
function) with average transverse flow velocity is used to
describe the contribution of participant or constituent quarks to
transverse momentum of considered hadron. The experimental spectra
of $\pi^{\mp}$ and $K^{\mp}$ (or $\overline{p}$ and $p$) are
fitted by the convolution of two (or three) TP-like functions due
to the fact that two (or three) constituent quarks are regarded as
two (or three) energy resources in the formation of considered
hadron. From the reasonable fits to the spectra, the thermal
freeze-out parameters are extracted, and the pseudo-entropy is
newly defined and extracted. Some parameters quickly change in the
energy range of less than 7.7 GeV, and slowly change in the energy
range of greater than 7.7 GeV, indicating the variation of
collision mechanism at around 7.7 GeV.
\\
\\
\end{abstract}

\maketitle

\section{Introduction}

As the basic gauge field theory~\cite{1,2,3} which is used for
describing strong interactions, Quantum Chromodynamics (QCD)
predicts that under the condition of high temperature and high
density~\cite{4}, partons can be released from the confined hadron
phase to form a new form of substance that is called Quark-Gluon
Plasma (QGP)~\cite{5,6,7,8}. The high-temperature and high-density
fireball formed in relativistic heavy-ion collisions can take the
shape of this kind of new form of strongly interacting substance.
In the meantime, the particular system composed by this kind of
substance is usually described by the QCD phase
diagram~\cite{9,10}. Nevertheless, once the external conditions
are changed, the strongly interacting substances described by QCD
will undergo a phase transition~\cite{9,10}. On the one hand, if
the thermodynamic properties of the system are expressed by the
temperature and the chemical potential of baryons~\cite{11,12,13},
the first-order phase transition will occur in the region where
the chemical potential of baryons is higher and the temperature is
lower. On the other hand, in the region where both the baryon
chemical potential~\cite{11,12,13} and temperature are high, the
system maintains dynamic balance which passes a smooth transition.
In other words, there is a critical point~\cite{14,15,16,17,18,19}
from hadronic to QGP phases in the deconfinement process, that is
the end point of the first-order phase transition of the QCD
matter.

One of the most important targets of relativistic heavy-ion
collisions~\cite{20,21,22,23} is to explore the QCD phase diagram,
understand the structure of QGP, and determine the phase
boundary~\cite{24,25} between different phases. The Beam Energy
Scan (BES) project~\cite{26,27,28,29} started by the Relativistic
Heavy Ion Collider (RHIC) in 2010 is to study the phase
diagram~\cite{9,10} of strongly interacting nuclear matter. The
RHIC-BES program~\cite{30,31,32} is performed for mainly three
objectives: The first one is to find the onset energy of
deconfinement phase transition~\cite{9,10}. The second one is to
identify the critical point of the QCD phase
diagram~\cite{9,10,14,15,16,17}. And the third one is to determine
the characteristics of the first-order phase transition. Since the
baryon chemical potential of matter is related to the collision
energy~\cite{11,12,13}, researchers expect to find the
corresponding critical point of phase transition by the way of
changing the collision energy. The RHIC-BES program including its
fixed target experiments is able to vary the collision energy over
a wide range in low energy region, thereby achieving an extensive
range of baryon chemical potential, further expanding the search
objective.

Hadrons are particles involved in strong interactions, including
two kinds of particles: mesons (bosons) and baryons (fermions).
The most common mesons measured in experiments include, but are
not limited to, negatively and positively charged $\pi$ mesons
($\pi^-$ and $\pi^+$), negatively and positively charged $K$
mesons ($K^-$ and $K^+$), etc. The most common baryons measured in
experiments include, but are not limited to, anti-protons and
protons ($\overline{p}$ and $p$), etc. In the RHIC-BES program and
at the previous and lower energy such as the Alternating Gradient
Synchrotron (AGS) energy, $\pi^-$, $\pi^+$, $K^-$, $K^+$,
$\overline{p}$, and $p$ are particularly important. In fact, the
abundant transverse momentum (mass) spectra of the mentioned
charged hadrons can be used to extract the common thermal
parameters such as the freeze-out temperature and transverse flow
velocity. Meanwhile, the novel pseudo-entropy, which will be
defined later in this paper, can be also extracted from the
spectra.

In this paper, the transverse momentum spectra of $\pi^{-}$,
$\pi^{+}$, $K^{-}$, $K^{+}$, $\overline{p}$, and $p$ produced at
mid-rapidity (mid-$y$) in gold-gold (Au-Au) collisions with
different centralities at the RHIC and its BES
energy~\cite{33,34,35} are collected. Meanwhile, the transverse
mass spectra of the mentioned charged hadrons produced at mid-$y$
in central Au-Au collisions at the AGS energy~\cite{36,37,38} are
also collected, though the spectra of $\overline{p}$ are not
available due to low energy. These spectra are used to extract the
thermal freeze-out parameters and the pseudo-entropy.

The remainder of this paper is structured as follows. The
formalism and method are briefly introduced in Section 2. The
results and discussion are given in Section 3. In Section 4, we
summarize our main observations and conclusions.

\section{Formalism and method}

Non-extensive thermodynamics is a new method for studying
heavy-ion collisions at relativistic energy. In the collisions
such as in Au-Au collisions, the Tsallis--Pareto-type
function~\cite{39,40,41,42,43,43a} can fit the transverse momentum
($p_T$) spectra in low and intermediate regions, particularly in the
final-state or hadronization process, demonstrating a strong
relation among particles. However, in very low-$p_T$ region, the
fit result is not ideal, and the meaning of Tsallis parameters
remains an open question.

In order to better fit the spectra in very low-$p_T$ region, we
express the information contained in the parameter more
intuitively using the modified Tsallis--Pareto-type function (the
TP-like function) as follows~\cite{44,46}:
\begin{align}
f_{p_T}(p_T)=\frac{1}{N}\frac{dN}{dp_T} = C p_{T}^{a_{0}}
\bigg(1+\frac{m_{T}-m_{0}}{nT}\bigg)^{-n}.
\end{align}
Here $N$ is the number of particles, $C$ is the normalization
constant, $m_{0}$ is the rest mass of the considered particle, $n$
is the power index that describes the degree of non-equilibrium,
$T$ is the effective temperature of the collision system, $a_{0}$
is the correction index, and $m_{T} =\sqrt{p_{T}^{2}+m_{0}^{2}}$
is the transverse mass of the particle. As an extention, the
TP-like function is naturally converged to the
Tsallis--Pareto-type function if we set $a_0=1$ in Eq. (1).

In order to obtain the thermal or kinetic (or kinematic)
freeze-out temperature $T_{0}$ and the average transverse flow
velocity $\langle\beta_{t}\rangle$~\cite{53,54}, one may fit
firstly the $p_T$ spectra of particles to obtain the effective
temperature $T$. The average $p_T$ ($\langle p_T \rangle$) and
average energy ($\overline{m}$) of particles in the source rest
frame can be obtained by using the Monte Carlo algorithm. Then,
one may extract the intercept as $T_{0}$ in the linear relation of
$T$ versus $m_{0}$, and further obtain the slope as
$\langle\beta_{t}\rangle$ in the linear relation of $\langle
p_{T}\rangle$ versus $\overline{m}$.

For purpose of simplifying this complicated solution process,
based on the idea of other work~\cite{62}, we modify $m_{T}$ and
$p_T$ in Eq. (1). For clarity, we use $m'_T$ and $p'_T$ instead of
$m_T$ and $p_T$ in Eq. (1) respectively. The Lorentz-like
transformation is $m'_{T} =
\langle\gamma_{t}\rangle(m_{T}-p_{T}\langle \beta_{t}\rangle)$ and
$|p'_T| = \langle\gamma_t\rangle|p_T-m_T
\langle\beta_{t}\rangle|$, where
$\langle\gamma_{t}\rangle=1/\sqrt{1-\langle\beta_{t}\rangle ^{2}}$
is the Lorentz-like factor. The absolute value $|p_T-m_T
\langle\beta_{t}\rangle|$ is used due to the fact that $p'_T$ is
positive and $p_T-m_T \langle\beta_{t}\rangle$ is possibly
negative in low-$p_T$ region. After the conversion, the new
TP-like function $f(p_T)$ certainly obeys the relation
$f_{p'_T}(p'_T)|dp'_T|=f(p_T)|dp_T|$. We have
\begin{align}
\begin{split}
f(p_T) &= C \frac{\langle\gamma_t\rangle^{a_0+1}}{m_T}
\big(m_T-p_T\langle\beta_t\rangle \big)
\big|p_T-m_T \langle\beta_{t}\rangle \big|^{a_0}\\
&\quad \times \bigg[1+\frac{\langle\gamma_{t}\rangle
(m_{T}-p_{T}\langle\beta_{t}\rangle)-m_{0}}{nT_{0}}\bigg]^{-n}.
\end{split}
\end{align}
One can see that $T$ in Eq. (1) is naturally converted to $T_0$ in
Eq. (2).

Our exploratory research shows that although Eq. (2) is applicable
in the fit of $p_T$ spectra, it is not flexible in some cases and
explicable in depth at quark level due to $m_0$ being the rest
mass of the considered particle. Empirically, Eq. (2) can be regarded
as the probability density function obeyed by the transverse momentum
$p_{ti}$ of the $i$-th constituent quark that contributes to $p_T$
of particle. Concretely, we have new TP-like function to be
\begin{align}
\begin{split}
f_{i}(p_{ti}) & = C_{i} \frac{\langle\gamma_t\rangle^{a_0+1}}{m_{ti}}
\big(m_{ti}-p_{ti}\langle\beta_t\rangle \big)
\big|p_{ti}-m_{ti} \langle\beta_{t}\rangle \big|^{a_0}\\
&\quad \times\bigg[1+\frac{\langle\gamma_{t}\rangle(m_{ti}-p_{ti}
\langle\beta_{t}\rangle)-m_{0i}}{nT_{0}}\bigg]^{-n},
\end{split}
\end{align}
where $m_{0i}$ is the constituent mass (0.31 MeV$/c^2$ for $u$ and
$d$ quarks and 0.5 GeV$/c^2$ for $s$ quark, as given in general
textbook~\cite{65a}) and $m_{ti}=\sqrt{p_{ti}^2+m_{0i}^2}$ is the
transverse mass of the $i$-th quark.

For a meson, $p_T=p_{t1}+p_{t2}$ due to two constituent quarks
which contribute independently. The probability density function
of meson's $p_T$ is the convolution of two TP-like functions. For
a baryon, $p_T=p_{t1}+p_{t2}+p_{t3}$ due to three constituent
quarks which also contribute independently. The probability
density function of baryon's $p_T$ is the convolution of three
TP-like functions. This analysis is at the quark level due to the
fact that it is based on the constituent mass of quark. The
present method is similar to the analysis at the particle level in
terms of the similar convolution.

For mesons, we have the convolution of two TP-like functions to be
\begin{align}
\begin{split}
f(p_{T})&=\int_{0}^{p_{T}} f_{1}(p_{t1})f_{2}(p_{T}-p_{t1})dp_{t1}\\
&=\int_{0}^{p_{T}} f_{2}(p_{t2})f_{1}(p_{T}-p_{t2})dp_{t2}.
\end{split}
\end{align}
For baryons, we have the convolution of the first two
TP-like functions to be
\begin{align}
\begin{split}
f_{12}(p_{t12})&=\int_{0}^{p_{t12}} f_{1}(p_{t1})f_{2}(p_{t12}-p_{t1})dp_{t1}\\
&=\int_{0}^{p_{t12}} f_{2}(p_{t2})f_{1}(p_{t12}-p_{t2})dp_{t2}.
\end{split}
\end{align}
The convolution of three TP-like functions is
\begin{align}
\begin{split}
f(p_{T})&=\int_{0}^{p_{T}} f_{12}(p_{t12})f_{3}(p_{T}-p_{t12})dp_{t12}\\
&=\int_{0}^{p_{T}}f_{3}(p_{t3})f_{12}(p_{T}-p_{t3})dp_{t3}.
\end{split}
\end{align}

In some cases, the spectra in experiments are for $m_T$. We need
to convert the probability density function [$f(p_T)$] of $p_T$ to
that [$f_{m_T}(m_T)$] of $m_T$. In fact, we have the relation
$f(p_{T})|dp_{T}|=f_{m_T}(m_{T})|dm_{T}|$. Further, we have
\begin{align}
f_{m_{T}}(m_{T})=\frac{m_{T}}{\sqrt{m_{T}^{2}-m_{0}^{2}}}
f\Big(\sqrt{m_{T}^{2}-m_{0}^{2}}\Big).
\end{align}
For invariant $p_T$ and $m_T$ spectra which are used usually in
experiments, we have
\begin{align}
\frac{1}{2\pi p_T}\frac{d^2N}{dp_Tdy}=\frac{1}{2\pi
m_T}\frac{d^2N}{dm_Tdy}.
\end{align}
Other forms of experimental spectra can be transformed from Eq.
(8) conveniently if $dy$ and the cross section are considered
reasonably.

According to statistical physics and information theory, the
essence of entropy is an expression of the degree of chaos existed
inherently in a system~\cite{68,69,70,71}. Entropy has a certain
relationship with the probability density function in
thermodynamics. In order to further study the role of color
confinement in relativistic heavy-ion collisions, the quantum
entanglement entropy (the entropy of gluons or partons) is
proposed to explore the underlying mechanism. As we know, the
relationship between the measured particle multiplicity
distribution in the deep inelastic scattering process and the
entropy of final hadrons is expressed as~\cite{68,69,70,71}
\begin{align}
S_{hadron}=-\sum P(N)\ln P(N),
\end{align}
where $P(N)$ is the multiplicity distribution of the charged
particles measured in a given measurement area. The entropy of
hadrons is assumed to be equal to the partonic
entropy~\cite{68,69,70,71}.

The TP-like function and the corresponding convolutions, which are
used to describe the $p_{ti}$ and $p_T$ distributions
respectively, are also the probability density functions.
Analogous to the multiplicity distribution of the above formula,
for the first time, we define the pseudo-entropy as follows:
\begin{align}
S'_{hadron}=-\sum f(p_{T})\ln f(p_{T})
\end{align}
in which the $p_T$ distribution $f(p_T)$ replaced the multiplicity
distribution $P(N)$ in Eq. (9). Different from $P(N)$ in which $N$
is discrete, $p_T$ in $f(p_T)$ is continuous. To give a suitable
and uniform description, we set the bin width of $p_T$ to be 0.1
GeV/$c$ in the considered energy range in this paper. Although the
pseudo-entropy is the bin width dependent, we may compare the
relative sizes in different cases if the bin width is beforehand
fixed.

\section{Results and discussion}

\subsection{Comparison with data}

\begin{figure*}[htbp]
\begin{center}
\includegraphics[width=14.0cm]{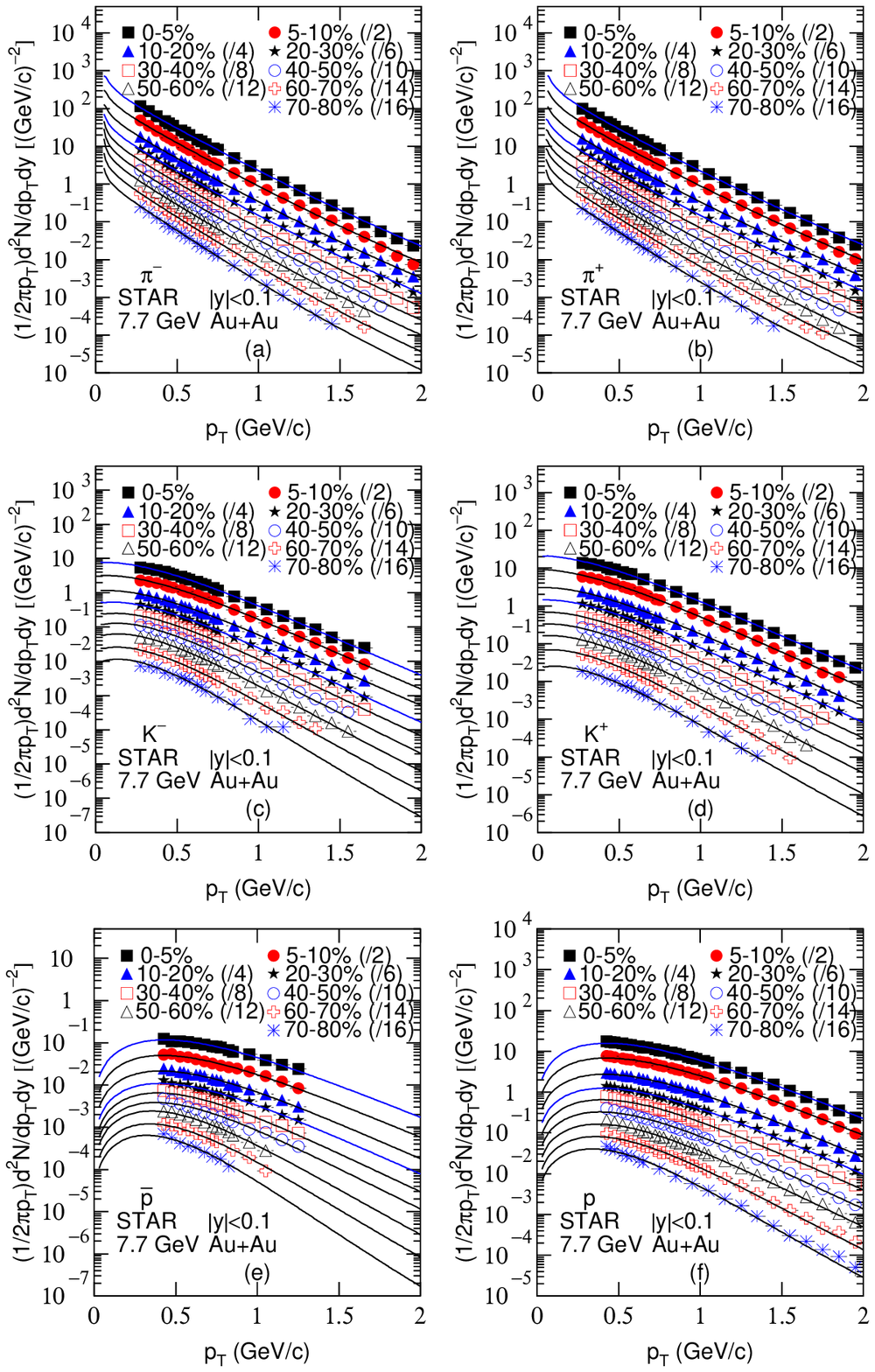}
\end{center}
\justifying\noindent {Figure 1. Transverse momentum spectra of
$\pi^{-}$ (a), $\pi^{+}$ (b), $K^{-}$ (c), $K^{+}$ (d),
$\overline{p}$ (e), and $p$ (f) produced in 7.7 GeV Au-Au
collisions with various centrality intervals and at the
mid-rapidity. The symbols represent the experimental data measured
by the STAR Collaboration in the RHIC-BES program~\cite{33,34} and
re-scaled by different amounts marked in the panels. The curves
are our results fitted by Eq. (4) for $\pi^{\mp}$ and $K^{\mp}$
[panels (a)--(d)] or Eq. (6) for $\overline{p}$ and $p$ [panels
(e) and (f)].}
\end{figure*}

\begin{table*} \vspace{-.5cm} \justifying\noindent {\small Table 1.
Values of $n$, $T_{0}$, $a_{0}$, $\langle\beta_{t}\rangle$,
$\chi^{2}$, and ndof corresponding to the curves in Figure 1.
\vspace{-0.5cm}

\begin{center}
\newcommand{\tabincell}[2]{\begin{tabular}{@{}#1@{}}#2\end{tabular}}
\begin{tabular} {cccccccccccc}\\ \hline\hline
Figure & Particle &  $\sqrt{s_{NN}}$ (GeV) & Selection & $n$ & $T_{0}$ (GeV) & $a_0$ & $\langle\beta_{t}\rangle$ ($c$) & $\chi^2$/ndof \\
\hline
Figure 1(a) & $\pi^{-}$  & $7.7$     & 0--5\%   & $17.8\pm0.7$ & $0.165\pm0.003$ & $-0.396\pm0.005$ & $0.112\pm0.003$ & $14/21$\\
        &                & $|y|<0.1$ & 5--10\%  & $16.5\pm0.7$ & $0.158\pm0.003$ & $-0.383\pm0.005$ & $0.109\pm0.003$ & $27/21$\\
        &                &           & 10--20\% & $15.0\pm0.5$ & $0.155\pm0.003$ & $-0.379\pm0.005$ & $0.106\pm0.003$ & $8/21$\\
        &                &           & 20--30\% & $14.9\pm0.5$ & $0.150\pm0.003$ & $-0.374\pm0.005$ & $0.102\pm0.003$ & $14/21$\\
        &                &           & 30--40\% & $14.6\pm0.5$ & $0.147\pm0.002$ & $-0.369\pm0.004$ & $0.099\pm0.002$ & $18/21$\\
        &                &           & 40--50\% & $14.4\pm0.5$ & $0.143\pm0.002$ & $-0.365\pm0.004$ & $0.098\pm0.002$ & $15/19$\\
        &                &           & 50--60\% & $14.0\pm0.4$ & $0.135\pm0.002$ & $-0.362\pm0.004$ & $0.098\pm0.002$ & $10/18$\\
        &                &           & 60--70\% & $13.8\pm0.4$ & $0.128\pm0.002$ & $-0.358\pm0.004$ & $0.094\pm0.002$ & $17/18$\\
        &                &           & 70--80\% & $13.7\pm0.4$ & $0.121\pm0.002$ & $-0.356\pm0.004$ & $0.093\pm0.002$ & $6/16$\\
\hline
Figure 1(b) & $\pi^{+}$  & $7.7$     & 0--5\%   & $18.2\pm0.8$ & $0.171\pm0.003$ & $-0.420\pm0.005$ & $0.119\pm0.003$ & $12/21$\\
        &                & $|y|<0.1$ & 5--10\%  & $17.3\pm0.7$ & $0.168\pm0.003$ & $-0.417\pm0.005$ & $0.116\pm0.003$ & $5/21$\\
        &                &           & 10--20\% & $16.0\pm0.6$ & $0.164\pm0.003$ & $-0.411\pm0.005$ & $0.114\pm0.003$ & $5/21$\\
        &                &           & 20--30\% & $15.7\pm0.6$ & $0.160\pm0.003$ & $-0.400\pm0.005$ & $0.111\pm0.003$ & $9/21$\\
        &                &           & 30--40\% & $15.3\pm0.6$ & $0.154\pm0.003$ & $-0.398\pm0.005$ & $0.109\pm0.003$ & $12/21$\\
        &                &           & 40--50\% & $14.9\pm0.5$ & $0.147\pm0.002$ & $-0.390\pm0.005$ & $0.106\pm0.003$ & $6/20$\\
        &                &           & 50--60\% & $14.4\pm0.5$ & $0.138\pm0.002$ & $-0.385\pm0.005$ & $0.104\pm0.003$ & $15/20$\\
        &                &           & 60--70\% & $13.9\pm0.4$ & $0.133\pm0.002$ & $-0.377\pm0.005$ & $0.102\pm0.003$ & $13/19$\\
        &                &           & 70--80\% & $13.4\pm0.4$ & $0.126\pm0.002$ & $-0.370\pm0.005$ & $0.100\pm0.002$ & $10/16$\\
\hline
Figure 1(c) & $K^{-}$    & $7.7$     & 0--5\%   & $27.9\pm1.7$ & $0.147\pm0.002$ & $0.011\pm0.004$  & $0.106\pm0.005$ & $9/18$\\
        &                & $|y|<0.1$ & 5--10\%  & $24.8\pm1.4$ & $0.143\pm0.002$ & $0.018\pm0.004$  & $0.105\pm0.005$ & $3/18$\\
        &                &           & 10--20\% & $22.8\pm1.2$ & $0.139\pm0.002$ & $0.025\pm0.004$  & $0.103\pm0.005$ & $5/18$\\
        &                &           & 20--30\% & $21.8\pm1.1$ & $0.129\pm0.002$ & $0.045\pm0.005$  & $0.102\pm0.005$ & $8/18$\\
        &                &           & 30--40\% & $21.5\pm1.1$ & $0.124\pm0.002$ & $0.061\pm0.005$  & $0.101\pm0.005$ & $5/18$\\
        &                &           & 40--50\% & $20.8\pm1.0$ & $0.115\pm0.002$ & $0.070\pm0.005$  & $0.096\pm0.004$ & $11/15$\\
        &                &           & 50--60\% & $20.6\pm1.0$ & $0.107\pm0.002$ & $0.081\pm0.005$  & $0.096\pm0.004$ & $5/16$\\
        &                &           & 60--70\% & $18.9\pm0.8$ & $0.100\pm0.002$ & $0.085\pm0.005$  & $0.094\pm0.004$ & $11/14$\\
        &                &           & 70--80\% & $18.2\pm0.8$ & $0.089\pm0.002$ & $0.093\pm0.005$  & $0.092\pm0.004$ & $26/11$\\
\hline
Figure 1(d) & $K^{+}$    & $7.7$     & 0--5\%   & $29.1\pm1.9$ & $0.164\pm0.003$ & $-0.022\pm0.004$ & $0.124\pm0.005$ & $9/18$\\
        &                & $|y|<0.1$ & 5--10\%  & $27.2\pm1.7$ & $0.160\pm0.003$ & $-0.020\pm0.004$ & $0.112\pm0.005$ & $6/20$\\
        &                &           & 10--20\% & $25.7\pm1.5$ & $0.155\pm0.003$ & $0.001\pm0.004$  & $0.109\pm0.005$ & $6/20$\\
        &                &           & 20--30\% & $25.1\pm1.5$ & $0.149\pm0.002$ & $0.005\pm0.004$  & $0.105\pm0.005$ & $5/20$\\
        &                &           & 30--40\% & $24.5\pm1.4$ & $0.144\pm0.002$ & $0.014\pm0.004$  & $0.099\pm0.004$ & $4/19$\\
        &                &           & 40--50\% & $22.5\pm1.2$ & $0.139\pm0.002$ & $0.019\pm0.004$  & $0.091\pm0.004$ & $9/18$\\
        &                &           & 50--60\% & $21.9\pm1.1$ & $0.129\pm0.002$ & $0.027\pm0.004$  & $0.085\pm0.004$ & $7/17$\\
        &                &           & 60--70\% & $20.1\pm1.0$ & $0.120\pm0.002$ & $0.034\pm0.005$  & $0.082\pm0.004$ & $8/16$\\
        &                &           & 70--80\% & $18.7\pm0.8$ & $0.112\pm0.002$ & $0.049\pm0.005$  & $0.079\pm0.004$ & $9/13$\\
\hline
Figure 1(e) & $\overline{p}$ & $7.7$ & 0--5\%   & $16.6\pm0.7$ & $0.128\pm0.002$ & $0.006\pm0.004$  & $0.347\pm0.008$ & $2/10$\\
        &                & $|y|<0.1$ & 5--10\%  & $15.8\pm0.6$ & $0.123\pm0.002$ & $0.007\pm0.004$  & $0.336\pm0.008$ & $4/9$\\
        &                &           & 10--20\% & $15.1\pm0.6$ & $0.116\pm0.002$ & $0.011\pm0.004$  & $0.325\pm0.008$ & $5/13$\\
        &                &           & 20--30\% & $14.6\pm0.5$ & $0.113\pm0.002$ & $0.015\pm0.004$  & $0.317\pm0.008$ & $5/11$\\
        &                &           & 30--40\% & $13.6\pm0.4$ & $0.104\pm0.002$ & $0.017\pm0.004$  & $0.309\pm0.008$ & $6/12$\\
        &                &           & 40--50\% & $12.8\pm0.3$ & $0.098\pm0.002$ & $0.020\pm0.004$  & $0.291\pm0.007$ & $11/9$\\
        &                &           & 50--60\% & $12.5\pm0.3$ & $0.086\pm0.001$ & $0.020\pm0.004$  & $0.276\pm0.007$ & $1/8$\\
        &                &           & 60--70\% & $11.9\pm0.3$ & $0.079\pm0.001$ & $0.023\pm0.004$  & $0.271\pm0.007$ & $2/6$\\
        &                &           & 70--80\% & $11.2\pm0.3$ & $0.069\pm0.001$ & $0.024\pm0.004$  & $0.260\pm0.007$ & $2/4$\\
\hline
Figure 1(f) & $p$        & $7.7$     & 0--5\%   & $18.4\pm0.8$ & $0.129\pm0.002$ & $0.006\pm0.004$  & $0.346\pm0.008$ & $5/24$\\
        &                & $|y|<0.1$ & 5--10\%  & $17.6\pm0.7$ & $0.127\pm0.002$ & $0.011\pm0.004$  & $0.334\pm0.008$ & $4/24$\\
        &                &           & 10--20\% & $16.7\pm0.7$ & $0.120\pm0.002$ & $0.016\pm0.004$  & $0.321\pm0.008$ & $4/24$\\
        &                &           & 20--30\% & $16.4\pm0.7$ & $0.117\pm0.002$ & $0.017\pm0.004$  & $0.314\pm0.008$ & $5/24$\\
        &                &           & 30--40\% & $15.0\pm0.5$ & $0.112\pm0.002$ & $0.019\pm0.004$  & $0.304\pm0.008$ & $9/23$\\
        &                &           & 40--50\% & $14.6\pm0.5$ & $0.105\pm0.002$ & $0.021\pm0.004$  & $0.298\pm0.007$ & $11/23$\\
        &                &           & 50--60\% & $13.4\pm0.4$ & $0.100\pm0.002$ & $0.023\pm0.004$  & $0.282\pm0.007$ & $16/22$\\
        &                &           & 60--70\% & $12.5\pm0.3$ & $0.091\pm0.001$ & $0.025\pm0.004$  & $0.271\pm0.007$ & $20/13$\\
        &                &           & 70--80\% & $11.3\pm0.3$ & $0.079\pm0.001$ & $0.028\pm0.004$  & $0.260\pm0.007$ & $25/16$\\
\hline
\end{tabular}%
\end{center}}
\end{table*}

Figure 1 shows the $p_T$ spectra, $(1/2\pi p_T)d^2N/dp_Tdy$, of
$\pi^{-}$ (a), $\pi^{+}$ (b), $K^{-}$ (c), $K^{+}$ (d),
$\overline{p}$ (e), and $p$ (f) produced at mid-$y$ ($|y|<0.1$) in
Au-Au collisions at center-of-mass energy $\sqrt{s_{NN}}=7.7$ GeV
(per nucleon pair) with different centrality intervals. The
symbols represent the experimental data measured by the STAR
Collaboration in the RHIC-BES project~\cite{33,34} and re-scaled
by different amounts marked in the panels for clarity. The curves
in Figures 1(a)--1(d) represent the results fitted by the
convolution of two TP-like functions, for the spectra of
$\pi^{\mp}$ and $K^{\mp}$. The curves for the spectra of
$\overline{p}$ (e) and $p$ (f) are the results fitted by the
convolution of three TP-like functions. We have used the method of
least squares to determine the values of parameters, and the
method of statistical simulation to determine the errors of
parameters. The values of $n$, $T_{0}$, $a_{0}$,
$\langle\beta_{t}\rangle$, $\chi^2$, and degrees of freedom (ndof)
are listed in Table 1. It can be seen that the convolution of two
(three) TP-like functions fits well the experimental spectra of
$\pi^{\mp}$ and $K^{\mp}$ ($\overline{p}$ and $p$) produced in 7.7
GeV Au-Au collisions with different centrality intervals in the
available $p_T$ range.

Similarly, the $p_T$ spectra of $\pi^{-}$ (a), $\pi^{+}$ (b),
$K^{-}$ (c), $K^{+}$ (d), $\overline{p}$ (e), and $p$ (f) produced
at mid-$y$ in Au-Au collisions with different centrality intervals
at $\sqrt{s_{NN}}=11.5$, 14.5, 19.6, 27, and 39 GeV are
demonstrated in Figures 2--6, respectively, where the data are
cited from the STAR Collaboration~\cite{33,34}. The values of
parameters, $\chi^2$, and ndof are listed in Tables 2--6 for
different energies. One can see that the convolution of two
(three) TP-like functions fits well the experimental spectra of
$\pi^{\mp}$ and $K^{\mp}$ ($\overline{p}$ and $p$) produced in
Au-Au collisions at 11.5, 14.5, 19.6, 27, and 39 GeV with
different centrality intervals in the available $p_T$ range.

In order to better explore the change trends of parameters,
besides the RHIC-BES energies, we have also conducted data
collection and research on the $p_T$ spectra at other RHIC
energies. Figures 7--9 show the $p_T$ spectra of $\pi^{-}$ (a),
$\pi^{+}$ (b), $K^{-}$ (c), $K^{+}$ (d), $\overline{p}$ (e), and
$p$ (f) produced in Au-Au collisions at 62.4, 130, and 200 GeV,
respectively. The data with $|y|<0.1$ are also cited from the STAR Collaboration~\cite{35}. The centrality intervals at 62.4 and 200
GeV are the same as those at RHIC-BES energies. However, the
centrality intervals at 130 GeV are somehow different from others.
The related parameters are listed in Tables 7--9 for different
energies. One can see again that the convolution of two (three)
TP-like functions fits well the experimental spectra of
$\pi^{\mp}$ and $K^{\mp}$ ($\overline{p}$ and $p$) produced in
Au-Au collisions at 62.4, 130, and 200 GeV with different
centrality intervals in the available $p_T$ range.

Except the above comparisons, we have studied the $m_T$
spectra of $\pi^{-}$ (a), $\pi^{+}$ (b), $K^{-}$ (c), $K^{+}$ (d),
and $p$ (e) produced at mid-$y$ in 0--5\% Au-Au collisions at
$\sqrt{s_{NN}}=2.70$, 3.32, 3.84, 4.30, and 4.88 GeV in Figure 10,
where the five energies are only available in panel (d). The
symbols for $\pi^{\mp}$ in $|y|<0.05$, $K^{\mp}$ in $|y|<0.25$,
and $p$ in $|y|<0.05$ represent the experimental data measured by
the E895~\cite{36}, E866~\cite{37}, and E895~\cite{38}
Collaborations, respectively, where some sets of data are
re-scaled by different amounts for clarity. The data for
$\overline{p}$ is not available in the same or similar experiments
due to low energy. The related parameters are listed in Table 10.
One can see that the convolution of two (three) TP-like functions
fits approximately the experimental spectra of $\pi^{\mp}$ and
$K^{\mp}$ ($p$) produced in 0--5\% Au-Au collisions at 2.70, 3.32,
3.84, 4.30, and 4.88 GeV.

\begin{figure*}[htbp]
\begin{center}
\includegraphics[width=14.0cm]{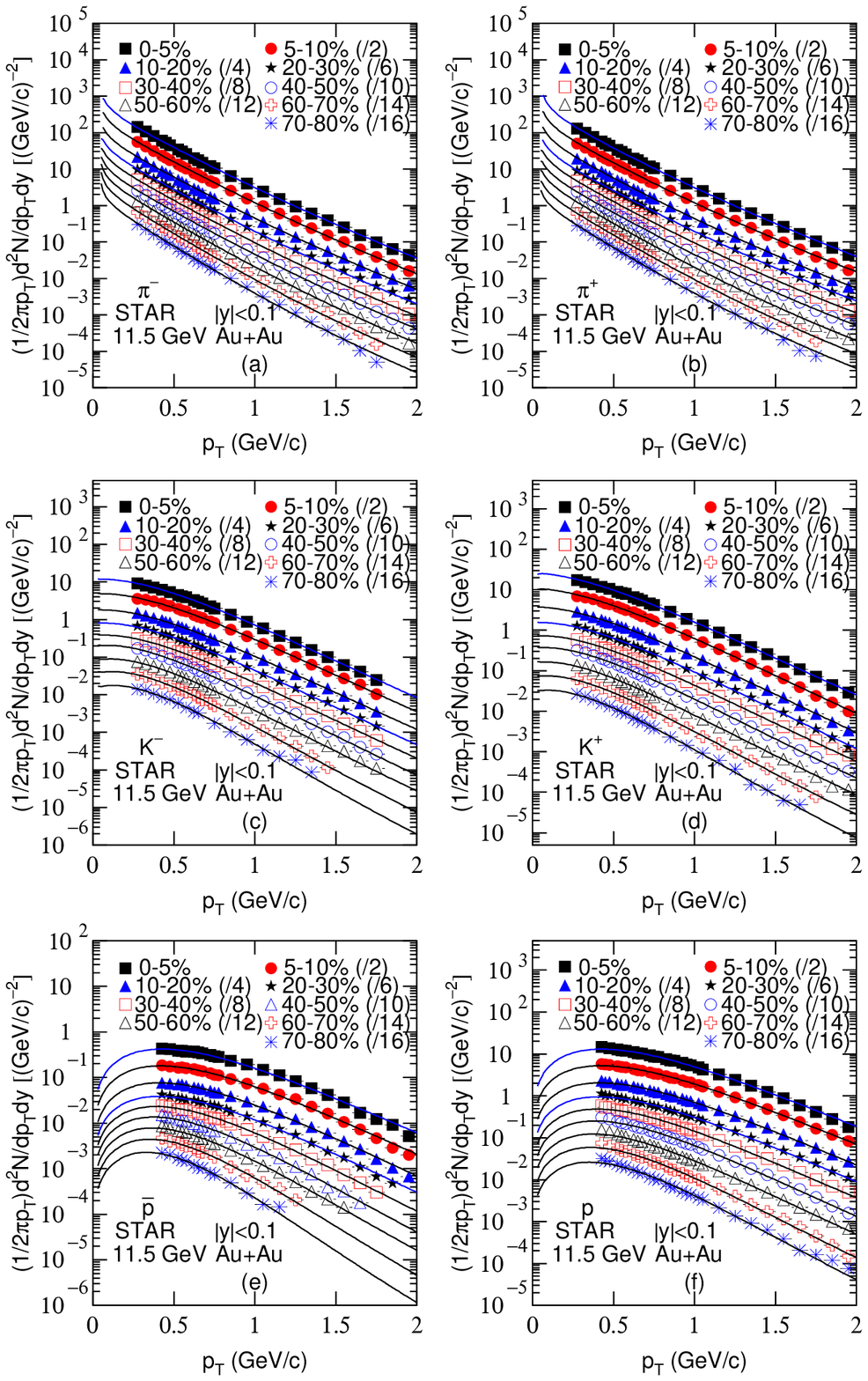}
\end{center}
\justifying\noindent {Figure 2. Same as Figure 1, but showing the
results for 11.5 GeV Au-Au collisions. The symbols represent the
STAR data~\cite{33}.}
\end{figure*}

\begin{table*} \vspace{-0.5cm} \justifying\noindent {\small Table 2.
Values of $n$, $T_{0}$, $a_{0}$, $\langle\beta_{t}\rangle$,
$\chi^{2}$, and ndof corresponding to the curves in Figure 2.
\vspace{-0.5cm}

\begin{center}
\newcommand{\tabincell}[2]{\begin{tabular}{@{}#1@{}}#2\end{tabular}}
\begin{tabular} {cccccccccccc}\\ \hline\hline
Figure & Particle &  $\sqrt{s_{NN}}$ (GeV) & Selection & $n$ & $T_{0}$ (GeV) & $a_0$ & $\langle\beta_{t}\rangle$ ($c$) & $\chi^2$/ndof \\
\hline
Figure 2(a) & $\pi^{-}$  & $11.5$    & 0--5\%   & $16.0\pm0.6$ & $0.168\pm0.003$ & $-0.397\pm0.005$ & $0.114\pm0.003$ & $3/21$\\
        &                & $|y|<0.1$ & 5--10\%  & $14.5\pm0.5$ & $0.162\pm0.003$ & $-0.396\pm0.005$ & $0.113\pm0.003$ & $6/21$\\
        &                &           & 10--20\% & $13.4\pm0.4$ & $0.158\pm0.003$ & $-0.389\pm0.005$ & $0.110\pm0.003$ & $6/21$\\
        &                &           & 20--30\% & $13.4\pm0.4$ & $0.153\pm0.003$ & $-0.382\pm0.005$ & $0.109\pm0.003$ & $20/21$\\
        &                &           & 30--40\% & $13.2\pm0.4$ & $0.150\pm0.003$ & $-0.381\pm0.005$ & $0.107\pm0.003$ & $11/21$\\
        &                &           & 40--50\% & $13.3\pm0.4$ & $0.147\pm0.002$ & $-0.376\pm0.005$ & $0.103\pm0.003$ & $15/21$\\
        &                &           & 50--60\% & $12.5\pm0.3$ & $0.138\pm0.002$ & $-0.368\pm0.005$ & $0.100\pm0.002$ & $28/21$\\
        &                &           & 60--70\% & $12.3\pm0.3$ & $0.133\pm0.002$ & $-0.370\pm0.005$ & $0.099\pm0.002$ & $28/19$\\
        &                &           & 70--80\% & $12.0\pm0.3$ & $0.128\pm0.002$ & $-0.365\pm0.004$ & $0.094\pm0.002$ & $29/19$\\
\hline
Figure 2(b) &$\pi^{+}$   & $11.5$    & 0--5\%   & $16.1\pm0.7$ & $0.175\pm0.003$ & $-0.422\pm0.005$ & $0.122\pm0.003$ & $3/21$\\
        &                & $|y|<0.1$ & 5--10\%  & $14.5\pm0.5$ & $0.170\pm0.003$ & $-0.420\pm0.005$ & $0.119\pm0.003$ & $5/21$\\
        &                &           & 10--20\% & $14.0\pm0.4$ & $0.166\pm0.003$ & $-0.417\pm0.005$ & $0.116\pm0.003$ & $5/21$\\
        &                &           & 20--30\% & $12.7\pm0.3$ & $0.161\pm0.003$ & $-0.407\pm0.005$ & $0.115\pm0.003$ & $4/21$\\
        &                &           & 30--40\% & $12.2\pm0.3$ & $0.156\pm0.003$ & $-0.400\pm0.005$ & $0.113\pm0.003$ & $3/21$\\
        &                &           & 40--50\% & $11.3\pm0.3$ & $0.149\pm0.002$ & $-0.391\pm0.005$ & $0.112\pm0.003$ & $6/21$\\
        &                &           & 50--60\% & $11.1\pm0.3$ & $0.140\pm0.002$ & $-0.386\pm0.005$ & $0.108\pm0.003$ & $10/21$\\
        &                &           & 60--70\% & $10.8\pm0.3$ & $0.135\pm0.002$ & $-0.378\pm0.005$ & $0.106\pm0.003$ & $8/19$\\
        &                &           & 70--80\% & $10.2\pm0.3$ & $0.129\pm0.002$ & $-0.374\pm0.005$ & $0.102\pm0.003$ & $20/19$\\
\hline
Figure 2(c) & $K^{-}$    & $11.5$    & 0--5\%   & $26.9\pm1.6$ & $0.153\pm0.003$ & $0.007\pm0.004$  & $0.114\pm0.005$ & $1/18$\\
        &                & $|y|<0.1$ & 5--10\%  & $24.3\pm1.4$ & $0.152\pm0.003$ & $0.011\pm0.004$  & $0.111\pm0.005$ & $5/19$\\
        &                &           & 10--20\% & $22.9\pm1.2$ & $0.149\pm0.002$ & $0.012\pm0.004$  & $0.109\pm0.005$ & $4/19$\\
        &                &           & 20--30\% & $21.8\pm1.1$ & $0.145\pm0.002$ & $0.015\pm0.004$  & $0.107\pm0.005$ & $4/19$\\
        &                &           & 30--40\% & $20.9\pm1.0$ & $0.141\pm0.002$ & $0.022\pm0.004$  & $0.107\pm0.005$ & $4/18$\\
        &                &           & 40--50\% & $17.5\pm0.7$ & $0.128\pm0.002$ & $0.032\pm0.005$  & $0.106\pm0.005$ & $11/18$\\
        &                &           & 50--60\% & $15.6\pm0.6$ & $0.121\pm0.002$ & $0.051\pm0.005$  & $0.103\pm0.005$ & $12/18$\\
        &                &           & 60--70\% & $15.1\pm0.6$ & $0.110\pm0.002$ & $0.064\pm0.005$  & $0.102\pm0.005$ & $8/15$\\
        &                &           & 70--80\% & $15.0\pm0.5$ & $0.103\pm0.002$ & $0.076\pm0.005$  & $0.092\pm0.004$ & $18/11$\\
\hline
Figure 2(d) & $K^{+}$    & $11.5$    & 0--5\%   & $28.3\pm1.8$ & $0.167\pm0.003$ & $-0.026\pm0.004$ & $0.127\pm0.005$ & $2/20$\\
        &                & $|y|<0.1$ & 5--10\%  & $27.4\pm1.7$ & $0.162\pm0.003$ & $-0.022\pm0.004$ & $0.121\pm0.005$ & $6/21$\\
        &                &           & 10--20\% & $24.9\pm1.4$ & $0.158\pm0.003$ & $-0.009\pm0.004$ & $0.115\pm0.005$ & $4/21$\\
        &                &           & 20--30\% & $23.1\pm1.3$ & $0.154\pm0.003$ & $0.003\pm0.004$  & $0.111\pm0.005$ & $8/21$\\
        &                &           & 30--40\% & $18.4\pm0.8$ & $0.149\pm0.002$ & $0.014\pm0.004$  & $0.108\pm0.005$ & $7/21$\\
        &                &           & 40--50\% & $17.0\pm0.7$ & $0.141\pm0.002$ & $0.019\pm0.004$  & $0.101\pm0.005$ & $16/21$\\
        &                &           & 50--60\% & $14.5\pm0.5$ & $0.134\pm0.002$ & $0.022\pm0.004$  & $0.099\pm0.004$ & $12/20$\\
        &                &           & 60--70\% & $13.9\pm0.4$ & $0.126\pm0.002$ & $0.032\pm0.005$  & $0.091\pm0.004$ & $5/18$\\
        &                &           & 70--80\% & $13.9\pm0.4$ & $0.116\pm0.002$ & $0.041\pm0.005$  & $0.089\pm0.004$ & $13/17$\\
\hline
Figure 2(e) & $\overline{p}$ & $11.5$& 0--5\%   & $16.1\pm0.7$ & $0.128\pm0.002$ & $0.003\pm0.004$  & $0.346\pm0.008$ & $19/18$\\
        &                & $|y|<0.1$ & 5--10\%  & $15.6\pm0.6$ & $0.124\pm0.002$ & $0.006\pm0.004$  & $0.330\pm0.008$ & $12/18$\\
        &                &           & 10--20\% & $15.1\pm0.6$ & $0.118\pm0.002$ & $0.010\pm0.004$  & $0.326\pm0.008$ & $13/18$\\
        &                &           & 20--30\% & $14.7\pm0.5$ & $0.116\pm0.002$ & $0.012\pm0.004$  & $0.314\pm0.008$ & $5/18$\\
        &                &           & 30--40\% & $14.0\pm0.4$ & $0.107\pm0.002$ & $0.015\pm0.004$  & $0.306\pm0.008$ & $3/18$\\
        &                &           & 40--50\% & $13.2\pm0.4$ & $0.098\pm0.002$ & $0.017\pm0.004$  & $0.296\pm0.007$ & $8/15$\\
        &                &           & 50--60\% & $12.1\pm0.3$ & $0.091\pm0.002$ & $0.020\pm0.004$  & $0.292\pm0.007$ & $7/14$\\
        &                &           & 60--70\% & $11.1\pm0.3$ & $0.081\pm0.001$ & $0.021\pm0.004$  & $0.284\pm0.007$ & $7/9$\\
        &                &           & 70--80\% & $10.1\pm0.3$ & $0.072\pm0.001$ & $0.022\pm0.004$  & $0.272\pm0.007$ & $6/9$\\
\hline
Figure 2(f) & $p$        & $11.5$    & 0--5\%   & $17.8\pm0.7$ & $0.129\pm0.002$ & $0.001\pm0.004$  & $0.347\pm0.008$ & $8/23$\\
        &                & $|y|<0.1$ & 5--10\%  & $16.2\pm0.7$ & $0.128\pm0.002$ & $0.004\pm0.004$  & $0.333\pm0.008$ & $2/24$\\
        &                &           & 10--20\% & $14.9\pm0.5$ & $0.122\pm0.002$ & $0.011\pm0.004$  & $0.327\pm0.008$ & $5/24$\\
        &                &           & 20--30\% & $13.8\pm0.4$ & $0.119\pm0.002$ & $0.013\pm0.004$  & $0.314\pm0.008$ & $7/24$\\
        &                &           & 30--40\% & $13.1\pm0.4$ & $0.115\pm0.002$ & $0.015\pm0.004$  & $0.302\pm0.008$ & $12/24$\\
        &                &           & 40--50\% & $12.4\pm0.3$ & $0.108\pm0.002$ & $0.017\pm0.004$  & $0.290\pm0.007$ & $11/23$\\
        &                &           & 50--60\% & $11.7\pm0.3$ & $0.104\pm0.002$ & $0.018\pm0.004$  & $0.274\pm0.007$ & $14/23$\\
        &                &           & 60--70\% & $11.1\pm0.3$ & $0.094\pm0.002$ & $0.021\pm0.004$  & $0.269\pm0.007$ & $18/23$\\
        &                &           & 70--80\% & $10.8\pm0.3$ & $0.088\pm0.001$ & $0.023\pm0.004$  & $0.265\pm0.007$ & $30/24$\\
\hline
\end{tabular}%
\end{center}}
\end{table*}

\begin{figure*}[!htb]
\begin{center}
\includegraphics[width=14.0cm]{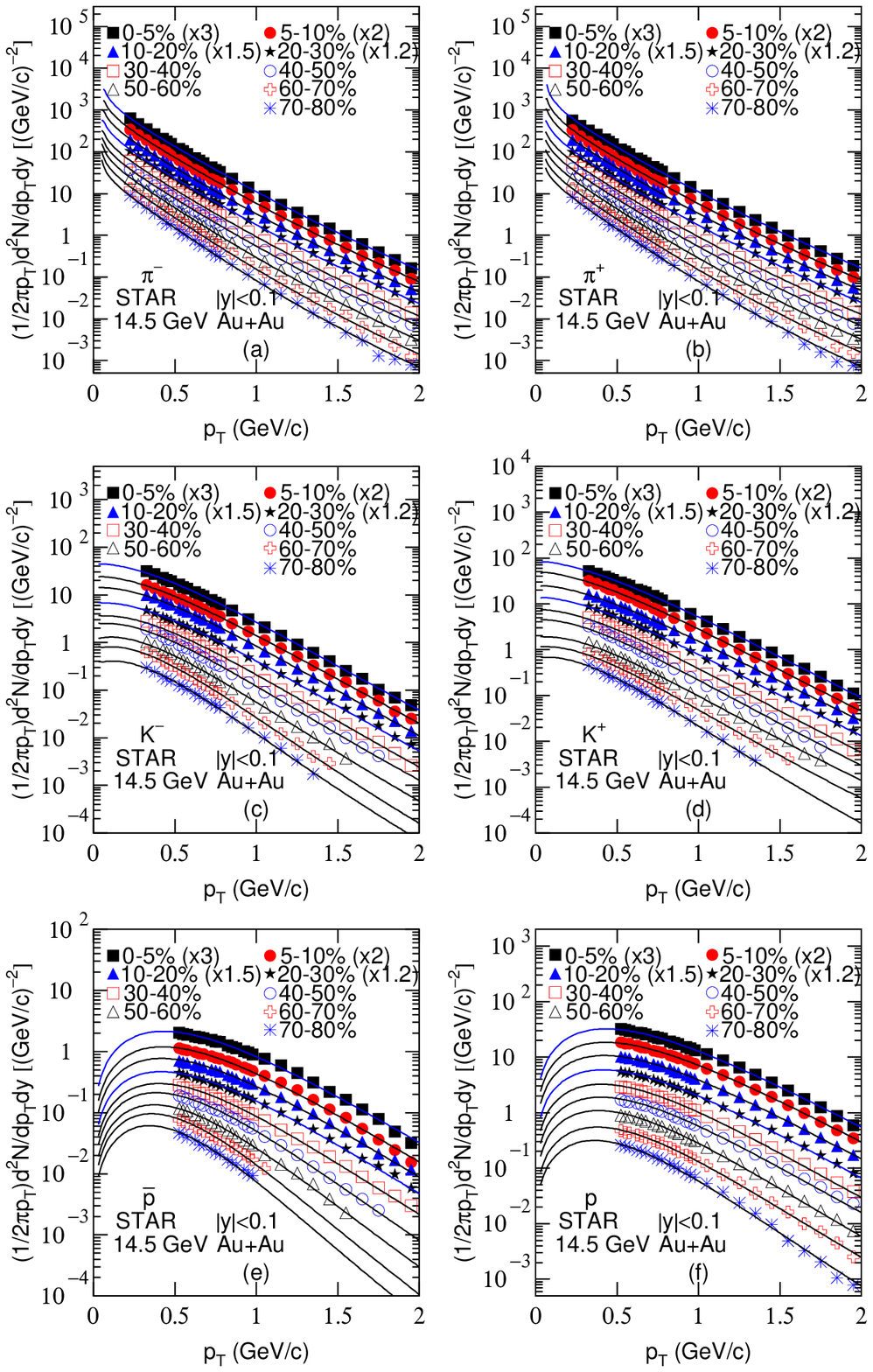}
\end{center}
\justifying\noindent {Figure 3. Same as Figure 1, but showing the
results for 14.5 GeV Au-Au collisions. The symbols represent the
STAR data~\cite{34}.}
\end{figure*}

\begin{table*} \vspace{-0.5cm} \justifying\noindent {\small Table 3.
Values of $n$, $T_{0}$, $a_{0}$, $\langle\beta_{t}\rangle$,
$\chi^{2}$, and ndof corresponding to the curves in Figure 3.
\vspace{-0.5cm}

\begin{center}
\newcommand{\tabincell}[2]{\begin{tabular}{@{}#1@{}}#2\end{tabular}}
\begin{tabular} {cccccccccccc}\\ \hline\hline
Figure & Particle &  $\sqrt{s_{NN}}$ (GeV) & Selection & $n$ & $T_{0}$ (GeV) & $a_0$ & $\langle\beta_{t}\rangle$ ($c$) & $\chi^2$/ndof \\
\hline
Figure 3(a) &$\pi^{-}$   & $14.5$    & 0--5\%   & $14.3\pm0.5$ & $0.171\pm0.003$ & $-0.407\pm0.005$ & $0.117\pm0.003$ & $1/23$\\
        &                & $|y|<0.1$ & 5--10\%  & $12.6\pm0.3$ & $0.165\pm0.003$ & $-0.405\pm0.005$ & $0.114\pm0.003$ & $2/23$\\
        &                &           & 10--20\% & $11.9\pm0.3$ & $0.161\pm0.003$ & $-0.396\pm0.005$ & $0.112\pm0.003$ & $3/23$\\
        &                &           & 20--30\% & $11.5\pm0.3$ & $0.158\pm0.003$ & $-0.392\pm0.005$ & $0.110\pm0.003$ & $4/23$\\
        &                &           & 30--40\% & $11.0\pm0.3$ & $0.153\pm0.003$ & $-0.385\pm0.005$ & $0.108\pm0.003$ & $4/23$\\
        &                &           & 40--50\% & $10.8\pm0.3$ & $0.150\pm0.003$ & $-0.378\pm0.005$ & $0.104\pm0.003$ & $3/23$\\
        &                &           & 50--60\% & $10.7\pm0.3$ & $0.141\pm0.002$ & $-0.377\pm0.005$ & $0.102\pm0.003$ & $13/23$\\
        &                &           & 60--70\% & $10.4\pm0.3$ & $0.133\pm0.002$ & $-0.370\pm0.005$ & $0.101\pm0.003$ & $25/23$\\
        &                &           & 70--80\% & $10.3\pm0.3$ & $0.128\pm0.002$ & $-0.368\pm0.004$ & $0.098\pm0.002$ & $25/23$\\
\hline
Figure 3(b) &$\pi^{+}$   & $14.5$    & 0--5\%   & $14.6\pm0.5$ & $0.179\pm0.003$ & $-0.424\pm0.005$ & $0.124\pm0.003$ & $5/23$\\
        &                & $|y|<0.1$ & 5--10\%  & $12.7\pm0.3$ & $0.172\pm0.003$ & $-0.421\pm0.005$ & $0.122\pm0.003$ & $2/23$\\
        &                &           & 10--20\% & $11.9\pm0.3$ & $0.167\pm0.003$ & $-0.419\pm0.005$ & $0.120\pm0.003$ & $4/23$\\
        &                &           & 20--30\% & $11.4\pm0.3$ & $0.163\pm0.003$ & $-0.416\pm0.005$ & $0.116\pm0.003$ & $4/23$\\
        &                &           & 30--40\% & $10.9\pm0.3$ & $0.159\pm0.003$ & $-0.411\pm0.005$ & $0.114\pm0.003$ & $3/23$\\
        &                &           & 40--50\% & $10.4\pm0.3$ & $0.151\pm0.003$ & $-0.404\pm0.005$ & $0.112\pm0.003$ & $1/23$\\
        &                &           & 50--60\% & $10.2\pm0.3$ & $0.143\pm0.002$ & $-0.396\pm0.005$ & $0.110\pm0.003$ & $7/23$\\
        &                &           & 60--70\% & $10.0\pm0.3$ & $0.137\pm0.002$ & $-0.394\pm0.005$ & $0.108\pm0.003$ & $14/23$\\
        &                &           & 70--80\% & $9.9\pm0.2$  & $0.129\pm0.002$ & $-0.384\pm0.005$ & $0.105\pm0.003$ & $26/23$\\
\hline
Figure 3(c) & $K^{-}$    & $14.5$    & 0--5\%   & $23.3\pm1.3$ & $0.159\pm0.003$ & $-0.015\pm0.004$ & $0.116\pm0.005$ & $2/21$\\
        &                & $|y|<0.1$ & 5--10\%  & $20.8\pm1.0$ & $0.157\pm0.003$ & $-0.013\pm0.004$ & $0.113\pm0.005$ & $2/21$\\
        &                &           & 10--20\% & $20.3\pm1.0$ & $0.152\pm0.003$ & $0.002\pm0.004$  & $0.112\pm0.005$ & $2/21$\\
        &                &           & 20--30\% & $18.9\pm0.8$ & $0.148\pm0.002$ & $0.014\pm0.004$  & $0.113\pm0.005$ & $7/21$\\
        &                &           & 30--40\% & $18.4\pm0.8$ & $0.143\pm0.002$ & $0.026\pm0.004$  & $0.107\pm0.005$ & $6/21$\\
        &                &           & 40--50\% & $13.7\pm0.4$ & $0.130\pm0.002$ & $0.029\pm0.004$  & $0.105\pm0.005$ & $5/19$\\
        &                &           & 50--60\% & $14.0\pm0.4$ & $0.122\pm0.002$ & $0.044\pm0.005$  & $0.102\pm0.005$ & $6/17$\\
        &                &           & 60--70\% & $13.2\pm0.3$ & $0.109\pm0.002$ & $0.053\pm0.005$  & $0.101\pm0.005$ & $6/16$\\
        &                &           & 70--80\% & $12.1\pm0.3$ & $0.104\pm0.002$ & $0.062\pm0.005$  & $0.098\pm0.004$ & $3/15$\\
\hline
Figure 3(d) & $K^{+}$    & $14.5$    & 0--5\%   & $26.6\pm1.6$ & $0.170\pm0.003$ & $-0.037\pm0.005$ & $0.129\pm0.005$ & $0.9/21$\\
        &                & $|y|<0.1$ & 5--10\%  & $22.1\pm1.2$ & $0.163\pm0.003$ & $-0.034\pm0.005$ & $0.125\pm0.005$ & $3/21$\\
        &                &           & 10--20\% & $17.8\pm0.7$ & $0.162\pm0.003$ & $-0.024\pm0.004$ & $0.123\pm0.005$ & $1/21$\\
        &                &           & 20--30\% & $16.7\pm0.7$ & $0.158\pm0.003$ & $-0.020\pm0.004$ & $0.112\pm0.005$ & $2/21$\\
        &                &           & 30--40\% & $15.9\pm0.6$ & $0.151\pm0.003$ & $-0.017\pm0.004$ & $0.108\pm0.005$ & $5/21$\\
        &                &           & 40--50\% & $13.6\pm0.4$ & $0.142\pm0.002$ & $-0.008\pm0.004$ & $0.105\pm0.005$ & $10/19$\\
        &                &           & 50--60\% & $11.8\pm0.3$ & $0.137\pm0.002$ & $0.012\pm0.004$  & $0.101\pm0.005$ & $6/19$\\
        &                &           & 60--70\% & $11.3\pm0.3$ & $0.127\pm0.002$ & $0.015\pm0.004$  & $0.095\pm0.004$ & $4/17$\\
        &                &           & 70--80\% & $12.3\pm0.3$ & $0.115\pm0.002$ & $0.025\pm0.004$  & $0.091\pm0.004$ & $2/15$\\
\hline
Figure 3(e) & $\overline{p}$ & $14.5$& 0--5\%   & $15.9\pm0.6$ & $0.127\pm0.002$ & $0.001\pm0.004$  & $0.358\pm0.008$ & $8/20$\\
        &                & $|y|<0.1$ & 5--10\%  & $14.6\pm0.5$ & $0.124\pm0.002$ & $0.004\pm0.004$  & $0.344\pm0.008$ & $9/20$\\
        &                &           & 10--20\% & $13.7\pm0.4$ & $0.120\pm0.002$ & $0.008\pm0.004$  & $0.339\pm0.008$ & $7/20$\\
        &                &           & 20--30\% & $13.3\pm0.4$ & $0.117\pm0.002$ & $0.011\pm0.004$  & $0.333\pm0.008$ & $5/20$\\
        &                &           & 30--40\% & $11.5\pm0.3$ & $0.108\pm0.002$ & $0.014\pm0.004$  & $0.324\pm0.008$ & $5/20$\\
        &                &           & 40--50\% & $11.7\pm0.3$ & $0.097\pm0.002$ & $0.016\pm0.004$  & $0.318\pm0.008$ & $2/18$\\
        &                &           & 50--60\% & $11.7\pm0.3$ & $0.089\pm0.001$ & $0.018\pm0.004$  & $0.312\pm0.008$ & $4/16$\\
        &                &           & 60--70\% & $10.5\pm0.3$ & $0.078\pm0.001$ & $0.020\pm0.004$  & $0.306\pm0.008$ & $3/11$\\
        &                &           & 70--80\% & $9.9\pm0.2$  & $0.073\pm0.001$ & $0.021\pm0.004$  & $0.298\pm0.007$ & $3/10$\\
\hline
Figure 3(f) & $p$        & $14.5$    & 0--5\%   & $17.0\pm0.7$ & $0.133\pm0.002$ & $-0.001\pm0.004$ & $0.350\pm0.008$ & $1/20$\\
        &                & $|y|<0.1$ & 5--10\%  & $15.4\pm0.6$ & $0.131\pm0.002$ & $0.001\pm0.004$  & $0.343\pm0.008$ & $0.9/20$\\
        &                &           & 10--20\% & $13.8\pm0.4$ & $0.126\pm0.002$ & $0.008\pm0.004$  & $0.331\pm0.008$ & $0.9/20$\\
        &                &           & 20--30\% & $13.1\pm0.4$ & $0.122\pm0.002$ & $0.011\pm0.004$  & $0.318\pm0.008$ & $2/20$\\
        &                &           & 30--40\% & $12.6\pm0.3$ & $0.118\pm0.002$ & $0.014\pm0.004$  & $0.304\pm0.008$ & $3/20$\\
        &                &           & 40--50\% & $12.1\pm0.3$ & $0.116\pm0.002$ & $0.016\pm0.004$  & $0.298\pm0.007$ & $7/20$\\
        &                &           & 50--60\% & $11.5\pm0.3$ & $0.107\pm0.002$ & $0.017\pm0.004$  & $0.290\pm0.007$ & $7/20$\\
        &                &           & 60--70\% & $10.9\pm0.3$ & $0.103\pm0.002$ & $0.020\pm0.004$  & $0.283\pm0.007$ & $4/20$\\
        &                &           & 70--80\% & $10.6\pm0.3$ & $0.093\pm0.002$ & $0.021\pm0.004$  & $0.272\pm0.007$ & $19/20$\\
\hline
\end{tabular}%
\end{center}}
\end{table*}

\begin{figure*}[!htb]
\begin{center}
\includegraphics[width=14.0cm]{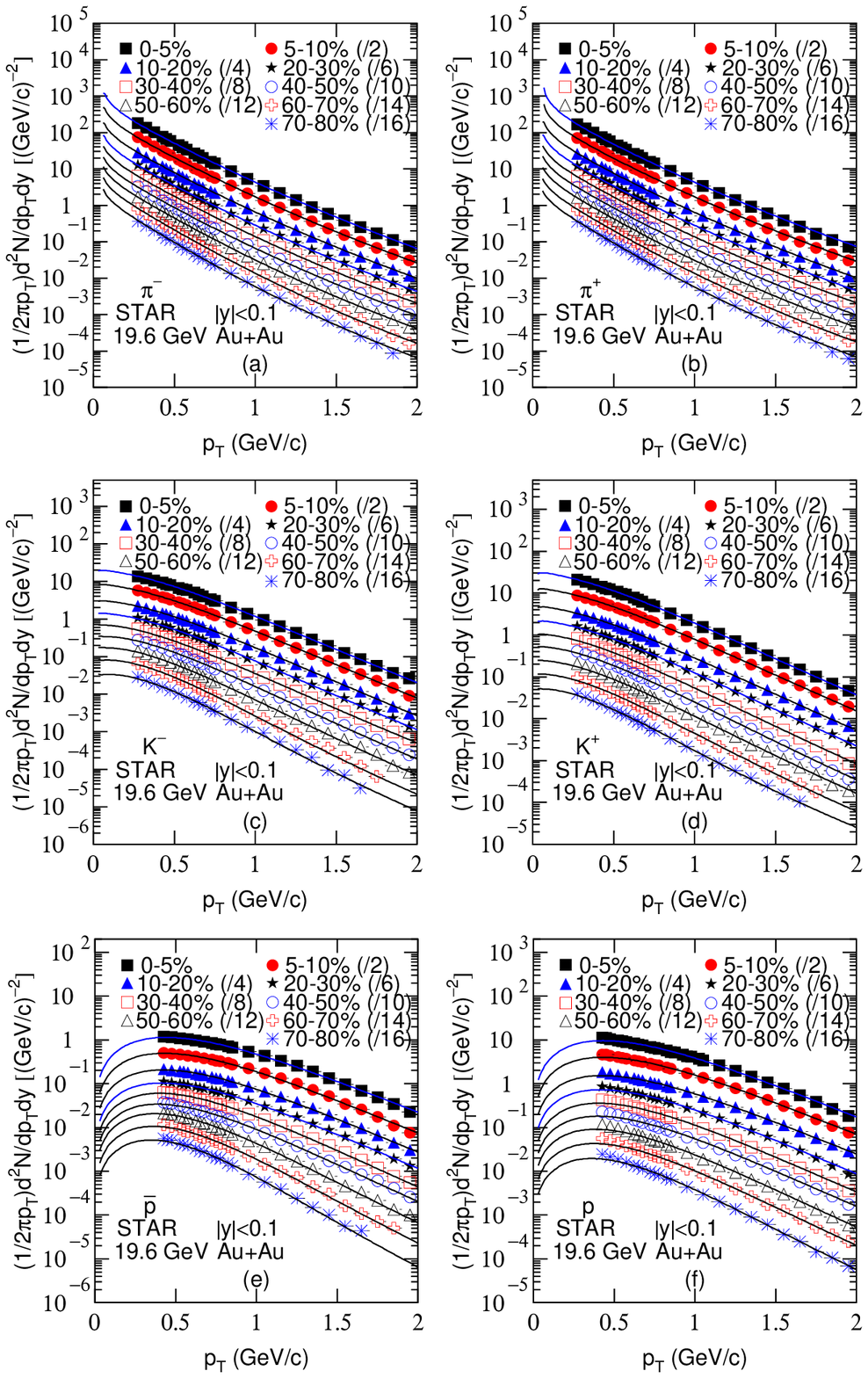}
\end{center}
\justifying\noindent {Figure 4. Same as Figure 1, but showing the
results for 19.6 GeV Au-Au collisions. The symbols represent the
STAR data~\cite{33}.}
\end{figure*}

\begin{table*} \vspace{-0.5cm} \justifying\noindent {\small Table 4.
Values of $n$, $T_{0}$, $a_{0}$, $\langle\beta_{t}\rangle$,
$\chi^{2}$, and ndof corresponding to the curves in Figure 4.
\vspace{-0.5cm}

\begin{center}
\newcommand{\tabincell}[2]{\begin{tabular}{@{}#1@{}}#2\end{tabular}}
\begin{tabular} {cccccccccccc}\\ \hline\hline
Figure & Particle &  $\sqrt{s_{NN}}$ (GeV) & Selection & $n$ & $T_0$ (GeV) & $a_0$ & $\langle\beta_{t}\rangle$ ($c$) & $\chi^2$/ndof \\
\hline
Figure 4(a) & $\pi^{-}$  & $19.6$    & 0--5\%   & $12.9\pm0.3$ & $0.173\pm0.003$ & $-0.410\pm0.005$ & $0.119\pm0.003$ & $1/21$\\
        &                & $|y|<0.1$ & 5--10\%  & $11.5\pm0.3$ & $0.168\pm0.003$ & $-0.409\pm0.005$ & $0.117\pm0.003$ & $2/21$\\
        &                &           & 10--20\% & $10.9\pm0.3$ & $0.164\pm0.003$ & $-0.402\pm0.005$ & $0.115\pm0.003$ & $2/21$\\
        &                &           & 20--30\% & $10.3\pm0.3$ & $0.161\pm0.003$ & $-0.404\pm0.005$ & $0.112\pm0.003$ & $2/21$\\
        &                &           & 30--40\% & $10.0\pm0.3$ & $0.157\pm0.003$ & $-0.395\pm0.005$ & $0.109\pm0.003$ & $3/21$\\
        &                &           & 40--50\% & $9.8\pm0.2$  & $0.153\pm0.003$ & $-0.390\pm0.005$ & $0.107\pm0.003$ & $3/21$\\
        &                &           & 50--60\% & $9.4\pm0.2$  & $0.145\pm0.002$ & $-0.383\pm0.005$ & $0.104\pm0.003$ & $4/21$\\
        &                &           & 60--70\% & $9.0\pm0.2$  & $0.136\pm0.002$ & $-0.377\pm0.005$ & $0.103\pm0.003$ & $13/21$\\
        &                &           & 70--80\% & $8.9\pm0.2$  & $0.131\pm0.002$ & $-0.372\pm0.005$ & $0.102\pm0.003$ & $13/20$\\
\hline
Figure 4(b) & $\pi^{+}$  & $19.6$    & 0--5\%   & $12.9\pm0.3$ & $0.183\pm0.003$ & $-0.424\pm0.005$ & $0.125\pm0.003$ & $14/21$\\
        &                & $|y|<0.1$ & 5--10\%  & $11.9\pm0.3$ & $0.177\pm0.003$ & $-0.420\pm0.005$ & $0.124\pm0.003$ & $6/21$\\
        &                &           & 10--20\% & $11.4\pm0.3$ & $0.170\pm0.003$ & $-0.417\pm0.005$ & $0.121\pm0.003$ & $1/21$\\
        &                &           & 20--30\% & $10.6\pm0.3$ & $0.166\pm0.003$ & $-0.415\pm0.005$ & $0.119\pm0.003$ & $1/21$\\
        &                &           & 30--40\% & $10.3\pm0.3$ & $0.162\pm0.003$ & $-0.414\pm0.005$ & $0.115\pm0.003$ & $2/21$\\
        &                &           & 40--50\% & $9.3\pm0.2$  & $0.153\pm0.003$ & $-0.409\pm0.005$ & $0.114\pm0.003$ & $3/21$\\
        &                &           & 50--60\% & $8.8\pm0.2$  & $0.146\pm0.002$ & $-0.401\pm0.005$ & $0.113\pm0.003$ & $5/21$\\
        &                &           & 60--70\% & $8.7\pm0.2$  & $0.139\pm0.002$ & $-0.395\pm0.005$ & $0.111\pm0.003$ & $4/21$\\
        &                &           & 70--80\% & $8.6\pm0.2$  & $0.132\pm0.002$ & $-0.387\pm0.005$ & $0.109\pm0.003$ & $18/21$\\
\hline
Figure 4(c) &$K^{-}$     & $19.6$    & 0--5\%   & $21.1\pm1.1$ & $0.165\pm0.003$ & $-0.034\pm0.005$ & $0.124\pm0.005$ & $2/21$\\
        &                & $|y|<0.1$ & 5--10\%  & $19.3\pm0.9$ & $0.161\pm0.003$ & $-0.030\pm0.005$ & $0.121\pm0.005$ & $2/21$\\
        &                &           & 10--20\% & $17.9\pm0.8$ & $0.155\pm0.003$ & $-0.018\pm0.004$ & $0.119\pm0.005$ & $3/21$\\
        &                &           & 20--30\% & $16.5\pm0.7$ & $0.151\pm0.003$ & $-0.016\pm0.004$ & $0.115\pm0.005$ & $7/21$\\
        &                &           & 30--40\% & $14.4\pm0.5$ & $0.144\pm0.002$ & $-0.009\pm0.004$ & $0.114\pm0.005$ & $12/21$\\
        &                &           & 40--50\% & $13.3\pm0.4$ & $0.133\pm0.002$ & $0.012\pm0.004$  & $0.106\pm0.005$ & $12/20$\\
        &                &           & 50--60\% & $13.1\pm0.4$ & $0.126\pm0.002$ & $0.016\pm0.004$  & $0.103\pm0.005$ & $18/20$\\
        &                &           & 60--70\% & $11.0\pm0.3$ & $0.111\pm0.002$ & $0.032\pm0.005$  & $0.102\pm0.005$ & $30/18$\\
        &                &           & 70--80\% & $10.3\pm0.3$ & $0.105\pm0.002$ & $0.044\pm0.005$  & $0.102\pm0.005$ & $25/16$\\
\hline
Figure 4(d) &$K^{+}$     & $19.6$    & 0--5\%   & $22.7\pm1.2$ & $0.172\pm0.003$ & $-0.045\pm0.005$ & $0.132\pm0.005$ & $1/21$\\
        &                & $|y|<0.1$ & 5--10\%  & $19.2\pm0.9$ & $0.168\pm0.003$ & $-0.034\pm0.005$ & $0.128\pm0.005$ & $3/21$\\
        &                &           & 10--20\% & $17.2\pm0.7$ & $0.164\pm0.003$ & $-0.030\pm0.005$ & $0.119\pm0.005$ & $3/21$\\
        &                &           & 20--30\% & $15.6\pm0.6$ & $0.159\pm0.003$ & $-0.026\pm0.004$ & $0.115\pm0.005$ & $5/21$\\
        &                &           & 30--40\% & $15.1\pm0.6$ & $0.151\pm0.003$ & $-0.020\pm0.004$ & $0.109\pm0.005$ & $15/21$\\
        &                &           & 40--50\% & $14.4\pm0.5$ & $0.145\pm0.002$ & $-0.017\pm0.004$ & $0.105\pm0.005$ & $14/20$\\
        &                &           & 50--60\% & $12.5\pm0.3$ & $0.141\pm0.002$ & $-0.011\pm0.004$ & $0.102\pm0.005$ & $23/20$\\
        &                &           & 60--70\% & $9.9\pm0.2$  & $0.127\pm0.002$ & $-0.006\pm0.004$ & $0.099\pm0.004$ & $15/18$\\
        &                &           & 70--80\% & $7.9\pm0.2$  & $0.115\pm0.002$ & $0.007\pm0.004$  & $0.093\pm0.004$ & $10/17$\\
\hline
Figure 4(e) & $\overline{p}$ & $19.6$& 0--5\%   & $15.3\pm0.6$ & $0.132\pm0.002$ & $0.001\pm0.004$  & $0.367\pm0.008$ & $5/17$\\
        &                & $|y|<0.1$ & 5--10\%  & $14.3\pm0.5$ & $0.126\pm0.002$ & $0.003\pm0.004$  & $0.353\pm0.008$ & $9/17$\\
        &                &           & 10--20\% & $13.2\pm0.4$ & $0.123\pm0.002$ & $0.006\pm0.004$  & $0.341\pm0.008$ & $5/19$\\
        &                &           & 20--30\% & $12.8\pm0.3$ & $0.120\pm0.002$ & $0.009\pm0.004$  & $0.328\pm0.008$ & $3/19$\\
        &                &           & 30--40\% & $12.4\pm0.3$ & $0.111\pm0.002$ & $0.013\pm0.004$  & $0.315\pm0.008$ & $5/20$\\
        &                &           & 40--50\% & $11.0\pm0.3$ & $0.106\pm0.002$ & $0.014\pm0.004$  & $0.305\pm0.008$ & $7/20$\\
        &                &           & 50--60\% & $10.2\pm0.3$ & $0.094\pm0.001$ & $0.015\pm0.004$  & $0.299\pm0.008$ & $11/20$\\
        &                &           & 60--70\% & $9.7\pm0.2$  & $0.086\pm0.001$ & $0.017\pm0.004$  & $0.301\pm0.008$ & $9/18$\\
        &                &           & 70--80\% & $9.5\pm0.2$  & $0.080\pm0.001$ & $0.019\pm0.004$  & $0.284\pm0.007$ & $11/17$\\
\hline
Figure 4(f) & $p$        & $19.6$    & 0--5\%   & $15.9\pm0.6$ & $0.133\pm0.002$ & $-0.002\pm0.004$ & $0.355\pm0.008$ & $7/24$\\
        &                & $|y|<0.1$ & 5--10\%  & $14.4\pm0.5$ & $0.132\pm0.002$ & $0.001\pm0.004$  & $0.345\pm0.008$ & $6/20$\\
        &                &           & 10--20\% & $13.3\pm0.4$ & $0.129\pm0.002$ & $0.005\pm0.004$  & $0.339\pm0.008$ & $5/18$\\
        &                &           & 20--30\% & $12.7\pm0.3$ & $0.122\pm0.002$ & $0.007\pm0.004$  & $0.328\pm0.008$ & $9/18$\\
        &                &           & 30--40\% & $12.3\pm0.3$ & $0.117\pm0.002$ & $0.010\pm0.004$  & $0.316\pm0.008$ & $8/18$\\
        &                &           & 40--50\% & $11.9\pm0.3$ & $0.116\pm0.002$ & $0.013\pm0.004$  & $0.305\pm0.008$ & $12/18$\\
        &                &           & 50--60\% & $11.3\pm0.3$ & $0.109\pm0.002$ & $0.015\pm0.004$  & $0.286\pm0.007$ & $15/18$\\
        &                &           & 60--70\% & $10.6\pm0.3$ & $0.104\pm0.002$ & $0.017\pm0.004$  & $0.280\pm0.007$ & $16/18$\\
        &                &           & 70--80\% & $9.8\pm0.2$  & $0.095\pm0.002$ & $0.019\pm0.004$  & $0.269\pm0.007$ & $13/18$\\
\hline
\end{tabular}%
\end{center}}
\end{table*}

\begin{figure*}[!htb]
\begin{center}
\includegraphics[width=14.0cm]{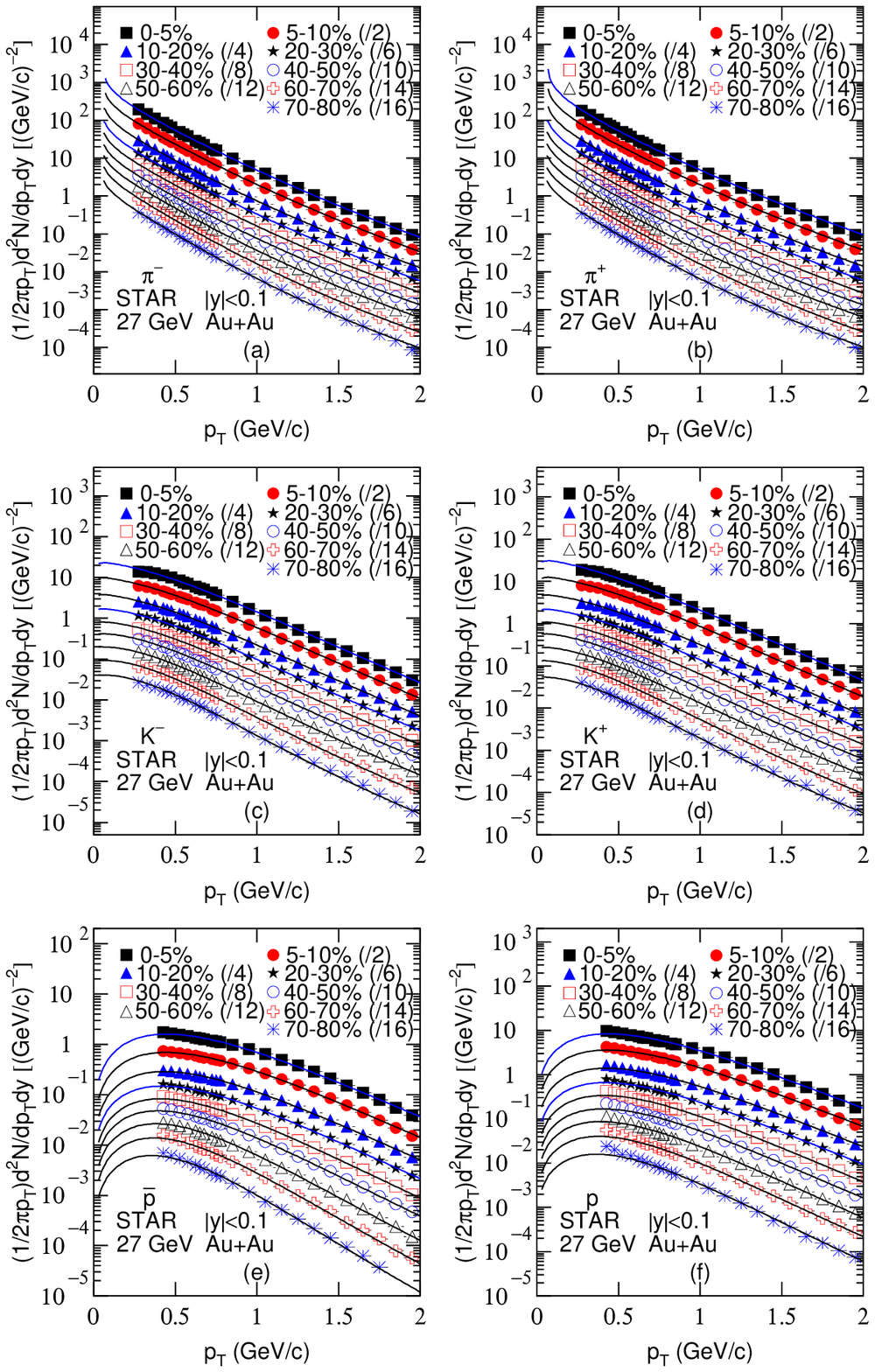}
\end{center}
\justifying\noindent {Figure 5. Same as Figure 1, but showing the
results for 27 GeV Au-Au collisions. The symbols represent the
STAR data~\cite{33}.}
\end{figure*}

\begin{table*} \vspace{-0.5cm} \justifying\noindent {\small Table 5.
Values of $n$, $T_{0}$, $a_{0}$, $\langle\beta_{t}\rangle$,
$\chi^{2}$, and ndof corresponding to the curves in Figure 5.
\vspace{-0.5cm}

\begin{center}
\newcommand{\tabincell}[2]{\begin{tabular}{@{}#1@{}}#2\end{tabular}}
\begin{tabular} {cccccccccccc}\\ \hline\hline
Figure & Particle &  $\sqrt{s_{NN}}$ (GeV) & Selection & $n$ & $T_0$ (GeV) & $a_0$ & $\langle\beta_{t}\rangle$ ($c$) & $\chi^2$/ndof \\
\hline
Figure 5(a) & $\pi^{-}$  & $27$      & 0--5\%   & $11.8\pm0.3$ & $0.175\pm0.003$ & $-0.410\pm0.005$ & $0.121\pm0.003$ & $4/21$\\
        &                & $|y|<0.1$ & 5--10\%  & $10.8\pm0.3$ & $0.170\pm0.003$ & $-0.409\pm0.005$ & $0.119\pm0.003$ & $1/21$\\
        &                &           & 10--20\% & $10.1\pm0.3$ & $0.167\pm0.003$ & $-0.407\pm0.005$ & $0.118\pm0.003$ & $2/21$\\
        &                &           & 20--30\% & $9.5\pm0.2$  & $0.164\pm0.003$ & $-0.405\pm0.005$ & $0.116\pm0.003$ & $1/21$\\
        &                &           & 30--40\% & $8.9\pm0.2$  & $0.160\pm0.003$ & $-0.403\pm0.005$ & $0.112\pm0.003$ & $1/21$\\
        &                &           & 40--50\% & $8.4\pm0.2$  & $0.155\pm0.003$ & $-0.398\pm0.005$ & $0.109\pm0.003$ & $3/21$\\
        &                &           & 50--60\% & $8.1\pm0.2$  & $0.148\pm0.002$ & $-0.388\pm0.005$ & $0.107\pm0.003$ & $2/21$\\
        &                &           & 60--70\% & $7.7\pm0.2$  & $0.139\pm0.002$ & $-0.385\pm0.005$ & $0.106\pm0.003$ & $4/21$\\
        &                &           & 70--80\% & $7.6\pm0.2$  & $0.133\pm0.002$ & $-0.373\pm0.005$ & $0.105\pm0.003$ & $13/21$\\
\hline
Figure 5(b) & $\pi^{+}$  & $27$      & 0--5\%   & $11.7\pm0.3$ & $0.186\pm0.003$ & $-0.425\pm0.005$ & $0.126\pm0.003$ & $14/21$\\
        &                & $|y|<0.1$ & 5--10\%  & $11.2\pm0.3$ & $0.180\pm0.003$ & $-0.424\pm0.005$ & $0.124\pm0.003$ & $6/21$\\
        &                &           & 10--20\% & $9.9\pm0.2$  & $0.172\pm0.003$ & $-0.421\pm0.005$ & $0.122\pm0.003$ & $3/21$\\
        &                &           & 20--30\% & $9.8\pm0.2$  & $0.168\pm0.003$ & $-0.416\pm0.005$ & $0.120\pm0.003$ & $1/21$\\
        &                &           & 30--40\% & $9.6\pm0.2$  & $0.164\pm0.003$ & $-0.415\pm0.005$ & $0.118\pm0.003$ & $1/21$\\
        &                &           & 40--50\% & $8.6\pm0.2$  & $0.156\pm0.003$ & $-0.409\pm0.005$ & $0.115\pm0.003$ & $1/21$\\
        &                &           & 50--60\% & $7.9\pm0.2$  & $0.149\pm0.002$ & $-0.402\pm0.005$ & $0.116\pm0.003$ & $4/21$\\
        &                &           & 60--70\% & $7.8\pm0.2$  & $0.142\pm0.002$ & $-0.396\pm0.005$ & $0.113\pm0.003$ & $4/21$\\
        &                &           & 70--80\% & $7.3\pm0.2$  & $0.135\pm0.002$ & $-0.395\pm0.005$ & $0.112\pm0.003$ & $9/21$\\
\hline
Figure 5(c) &$K^{-}$     & $27$      & 0--5\%   & $21.0\pm1.0$ & $0.166\pm0.003$ & $-0.038\pm0.005$ & $0.136\pm0.005$ & $12/20$\\
        &                & $|y|<0.1$ & 5--10\%  & $18.5\pm0.8$ & $0.164\pm0.003$ & $-0.036\pm0.005$ & $0.129\pm0.005$ & $4/21$\\
        &                &           & 10--20\% & $15.6\pm0.6$ & $0.158\pm0.003$ & $-0.028\pm0.004$ & $0.128\pm0.005$ & $4/21$\\
        &                &           & 20--30\% & $13.7\pm0.4$ & $0.152\pm0.003$ & $-0.019\pm0.004$ & $0.126\pm0.005$ & $5/21$\\
        &                &           & 30--40\% & $12.6\pm0.3$ & $0.148\pm0.002$ & $-0.010\pm0.004$ & $0.116\pm0.005$ & $5/21$\\
        &                &           & 40--50\% & $10.2\pm0.3$ & $0.137\pm0.002$ & $-0.002\pm0.004$ & $0.111\pm0.005$ & $6/21$\\
        &                &           & 50--60\% & $9.4\pm0.2$  & $0.128\pm0.002$ & $0.009\pm0.005$  & $0.110\pm0.005$ & $6/21$\\
        &                &           & 60--70\% & $8.9\pm0.2$  & $0.120\pm0.002$ & $0.014\pm0.004$  & $0.105\pm0.005$ & $10/21$\\
        &                &           & 70--80\% & $8.6\pm0.2$  & $0.111\pm0.002$ & $0.018\pm0.004$  & $0.105\pm0.005$ & $22/21$\\
\hline
Figure 5(d) & $K^{+}$    & $27$      & 0--5\%   & $21.9\pm1.1$ & $0.175\pm0.003$ & $-0.050\pm0.005$ & $0.144\pm0.005$ & $12/21$\\
        &                & $|y|<0.1$ & 5--10\%  & $19.2\pm0.9$ & $0.171\pm0.003$ & $-0.035\pm0.005$ & $0.137\pm0.005$ & $7/21$\\
        &                &           & 10--20\% & $17.2\pm0.7$ & $0.167\pm0.003$ & $-0.029\pm0.004$ & $0.133\pm0.005$ & $7/21$\\
        &                &           & 20--30\% & $13.4\pm0.4$ & $0.162\pm0.003$ & $-0.026\pm0.004$ & $0.126\pm0.005$ & $3/21$\\
        &                &           & 30--40\% & $11.0\pm0.3$ & $0.152\pm0.003$ & $-0.022\pm0.004$ & $0.123\pm0.005$ & $3/21$\\
        &                &           & 40--50\% & $9.5\pm0.2$  & $0.143\pm0.002$ & $-0.019\pm0.004$ & $0.115\pm0.005$ & $4/21$\\
        &                &           & 50--60\% & $9.1\pm0.2$  & $0.138\pm0.002$ & $-0.015\pm0.004$ & $0.109\pm0.005$ & $5/21$\\
        &                &           & 60--70\% & $8.6\pm0.2$  & $0.129\pm0.002$ & $-0.012\pm0.004$ & $0.102\pm0.005$ & $6/21$\\
        &                &           & 70--80\% & $7.5\pm0.2$  & $0.119\pm0.002$ & $-0.004\pm0.004$ & $0.096\pm0.004$ & $13/21$\\
\hline
Figure 5(e) & $\overline{p}$ & $27$  & 0--5\%   & $14.6\pm0.5$ & $0.134\pm0.002$ & $-0.001\pm0.004$ & $0.379\pm0.008$ & $5/17$\\
        &                & $|y|<0.1$ & 5--10\%  & $14.0\pm0.4$ & $0.130\pm0.002$ & $0.003\pm0.004$  & $0.375\pm0.008$ & $4/17$\\
        &                &           & 10--20\% & $12.4\pm0.3$ & $0.127\pm0.002$ & $0.005\pm0.004$  & $0.356\pm0.008$ & $3/17$\\
        &                &           & 20--30\% & $11.3\pm0.3$ & $0.124\pm0.002$ & $0.008\pm0.004$  & $0.330\pm0.008$ & $2/17$\\
        &                &           & 30--40\% & $10.7\pm0.3$ & $0.115\pm0.002$ & $0.011\pm0.004$  & $0.328\pm0.008$ & $3/17$\\
        &                &           & 40--50\% & $9.8\pm0.2$  & $0.110\pm0.002$ & $0.015\pm0.004$  & $0.304\pm0.008$ & $3/17$\\
        &                &           & 50--60\% & $9.2\pm0.2$  & $0.097\pm0.002$ & $0.016\pm0.004$  & $0.307\pm0.008$ & $7/17$\\
        &                &           & 60--70\% & $8.9\pm0.2$  & $0.090\pm0.002$ & $0.018\pm0.004$  & $0.300\pm0.007$ & $10/17$\\
        &                &           & 70--80\% & $8.8\pm0.2$  & $0.084\pm0.001$ & $0.019\pm0.004$  & $0.285\pm0.007$ & $10/15$\\
\hline
Figure 5(f) & $p$        & $27$      & 0--5\%   & $14.7\pm0.5$ & $0.136\pm0.002$ & $-0.005\pm0.004$ & $0.359\pm0.008$ & $6/18$\\
        &                & $|y|<0.1$ & 5--10\%  & $13.7\pm0.4$ & $0.133\pm0.002$ & $-0.003\pm0.004$ & $0.349\pm0.008$ & $6/18$\\
        &                &           & 10--20\% & $13.1\pm0.4$ & $0.130\pm0.002$ & $0.004\pm0.004$  & $0.339\pm0.008$ & $5/18$\\
        &                &           & 20--30\% & $11.6\pm0.3$ & $0.124\pm0.002$ & $0.005\pm0.004$  & $0.333\pm0.008$ & $6/18$\\
        &                &           & 30--40\% & $12.0\pm0.3$ & $0.119\pm0.002$ & $0.007\pm0.004$  & $0.328\pm0.008$ & $8/18$\\
        &                &           & 40--50\% & $12.1\pm0.3$ & $0.118\pm0.002$ & $0.011\pm0.004$  & $0.311\pm0.008$ & $13/18$\\
        &                &           & 50--60\% & $10.7\pm0.3$ & $0.109\pm0.002$ & $0.013\pm0.004$  & $0.298\pm0.007$ & $15/18$\\
        &                &           & 60--70\% & $10.0\pm0.3$ & $0.105\pm0.002$ & $0.016\pm0.004$  & $0.283\pm0.007$ & $20/18$\\
        &                &           & 70--80\% & $9.7\pm0.2$  & $0.100\pm0.002$ & $0.018\pm0.004$  & $0.264\pm0.007$ & $28/18$\\
\hline
\end{tabular}%
\end{center}}
\end{table*}

\begin{figure*}
\begin{center}
\includegraphics[width=14.0cm]{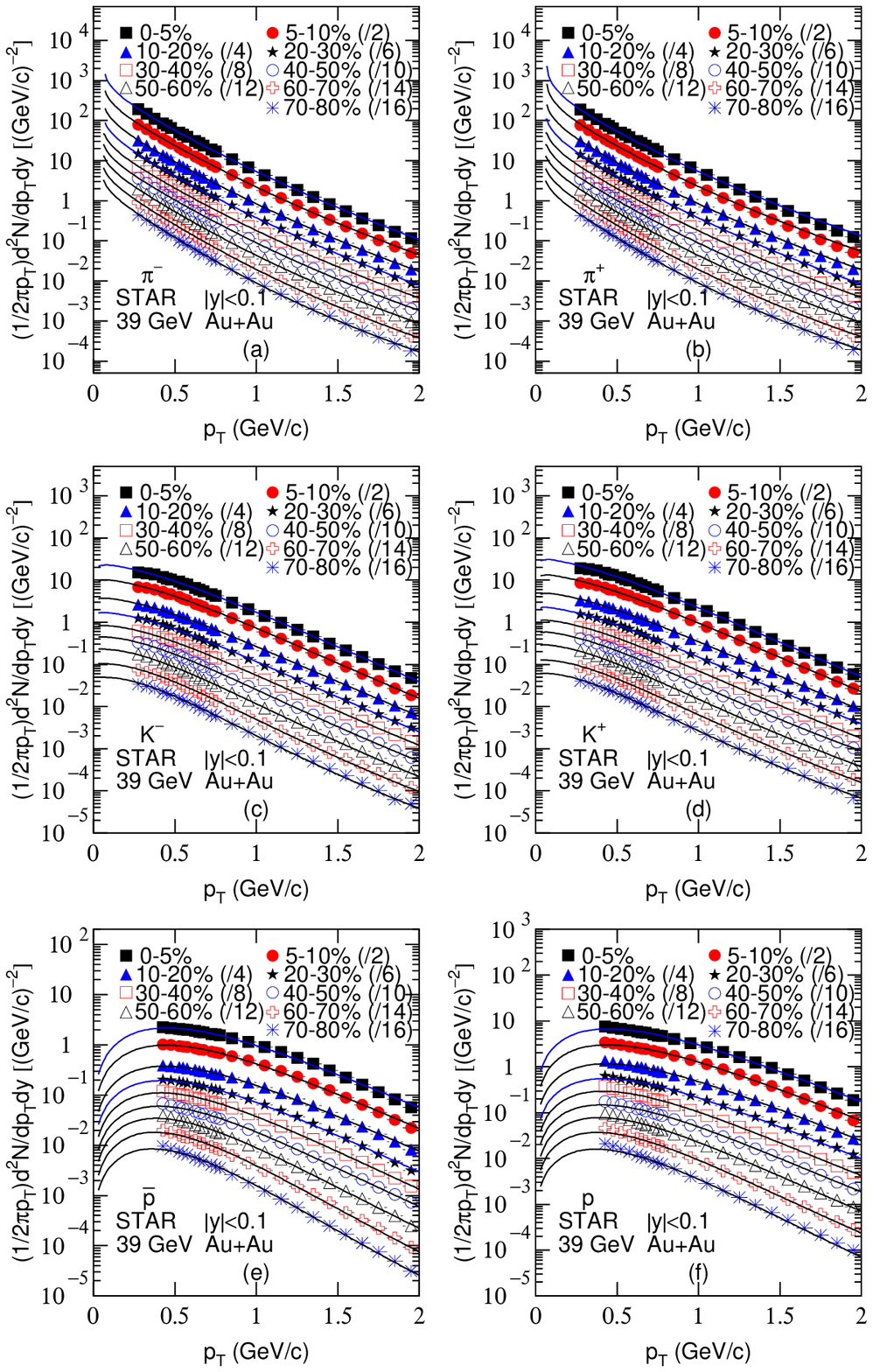}
\end{center}
\justifying\noindent {Figure 6. Same as Figure 1, but showing the
results for 39 GeV Au-Au collisions. The symbols represent the
STAR data~\cite{33}.}
\end{figure*}

\begin{table*} \vspace{-0.5cm} \justifying\noindent {\small Table 6.
Values of $n$, $T_{0}$, $a_{0}$, $\langle\beta_{t}\rangle$,
$\chi^{2}$, and ndof corresponding to the curves in Figure 6.
\vspace{-0.5cm}

\begin{center}
\newcommand{\tabincell}[2]{\begin{tabular}{@{}#1@{}}#2\end{tabular}}
\begin{tabular} {cccccccccccc}\\ \hline\hline
Figure & Particle &  $\sqrt{s_{NN}}$ (GeV) & Selection & $n$ & $T_0$ (GeV) & $a_0$ & $\langle\beta_{t}\rangle$ ($c$) & $\chi^2$/ndof \\
\hline
Figure 6(a) & $\pi^{-}$  & $39$      & 0--5\%   & $10.1\pm0.3$ & $0.179\pm0.003$ & $-0.418\pm0.005$ & $0.122\pm0.003$ & $2/21$\\
        &                & $|y|<0.1$ & 5--10\%  & $9.2\pm0.2$  & $0.173\pm0.003$ & $-0.414\pm0.005$ & $0.120\pm0.003$ & $2/21$\\
        &                &           & 10--20\% & $8.8\pm0.2$  & $0.170\pm0.003$ & $-0.412\pm0.005$ & $0.118\pm0.003$ & $1/21$\\
        &                &           & 20--30\% & $8.3\pm0.2$  & $0.167\pm0.003$ & $-0.411\pm0.005$ & $0.115\pm0.003$ & $1/21$\\
        &                &           & 30--40\% & $7.9\pm0.2$  & $0.162\pm0.003$ & $-0.403\pm0.005$ & $0.113\pm0.003$ & $1/21$\\
        &                &           & 40--50\% & $7.6\pm0.2$  & $0.158\pm0.003$ & $-0.399\pm0.005$ & $0.107\pm0.003$ & $1/21$\\
        &                &           & 50--60\% & $7.1\pm0.2$  & $0.150\pm0.003$ & $-0.394\pm0.005$ & $0.106\pm0.003$ & $1/21$\\
        &                &           & 60--70\% & $6.7\pm0.1$  & $0.142\pm0.002$ & $-0.391\pm0.005$ & $0.105\pm0.003$ & $3/21$\\
        &                &           & 70--80\% & $6.5\pm0.1$  & $0.136\pm0.002$ & $-0.384\pm0.005$ & $0.104\pm0.003$ & $6/21$\\
\hline
Figure 6(b) & $\pi^{+}$  & $39$      & 0--5\%   & $9.2\pm0.2$  & $0.190\pm0.003$ & $-0.428\pm0.005$ & $0.126\pm0.003$ & $28/21$\\
        &                & $|y|<0.1$ & 5--10\%  & $8.8\pm0.2$  & $0.183\pm0.003$ & $-0.426\pm0.005$ & $0.125\pm0.003$ & $19/21$\\
        &                &           & 10--20\% & $8.4\pm0.2$  & $0.175\pm0.003$ & $-0.422\pm0.005$ & $0.123\pm0.003$ & $5/21$\\
        &                &           & 20--30\% & $8.1\pm0.2$  & $0.170\pm0.003$ & $-0.418\pm0.005$ & $0.122\pm0.003$ & $2/21$\\
        &                &           & 30--40\% & $7.9\pm0.2$  & $0.166\pm0.003$ & $-0.417\pm0.005$ & $0.119\pm0.003$ & $1/21$\\
        &                &           & 40--50\% & $7.3\pm0.2$  & $0.158\pm0.003$ & $-0.411\pm0.005$ & $0.118\pm0.003$ & $2/21$\\
        &                &           & 50--60\% & $6.9\pm0.1$  & $0.151\pm0.003$ & $-0.406\pm0.005$ & $0.114\pm0.003$ & $1/21$\\
        &                &           & 60--70\% & $6.5\pm0.1$  & $0.144\pm0.002$ & $-0.403\pm0.005$ & $0.112\pm0.003$ & $3/21$\\
        &                &           & 70--80\% & $6.1\pm0.1$  & $0.137\pm0.002$ & $-0.399\pm0.005$ & $0.111\pm0.003$ & $5/21$\\
\hline
Figure 6(c) & $K^{-}$    & $39$      & 0--5\%   & $19.8\pm0.9$ & $0.171\pm0.003$ & $-0.039\pm0.005$ & $0.164\pm0.005$ & $6/21$\\
        &                & $|y|<0.1$ & 5--10\%  & $17.8\pm0.7$ & $0.167\pm0.003$ & $-0.033\pm0.005$ & $0.152\pm0.005$ & $5/21$\\
        &                &           & 10--20\% & $15.6\pm0.6$ & $0.165\pm0.003$ & $-0.024\pm0.004$ & $0.142\pm0.005$ & $3/21$\\
        &                &           & 20--30\% & $14.4\pm0.5$ & $0.161\pm0.003$ & $-0.019\pm0.004$ & $0.137\pm0.005$ & $5/21$\\
        &                &           & 30--40\% & $11.4\pm0.3$ & $0.152\pm0.003$ & $-0.009\pm0.004$ & $0.133\pm0.005$ & $5/21$\\
        &                &           & 40--50\% & $9.7\pm0.2$  & $0.139\pm0.002$ & $-0.005\pm0.004$ & $0.130\pm0.005$ & $7/21$\\
        &                &           & 50--60\% & $9.0\pm0.2$  & $0.132\pm0.002$ & $-0.002\pm0.004$ & $0.126\pm0.005$ & $9/21$\\
        &                &           & 60--70\% & $7.8\pm0.2$  & $0.124\pm0.002$ & $0.008\pm0.004$  & $0.123\pm0.005$ & $11/21$\\
        &                &           & 70--80\% & $7.0\pm0.1$  & $0.115\pm0.002$ & $0.016\pm0.004$  & $0.113\pm0.005$ & $15/21$\\
\hline
Figure 6(d) & $K^{+}$    & $39$      & 0--5\%   & $20.6\pm1.0$ & $0.177\pm0.003$ & $-0.051\pm0.005$ & $0.150\pm0.005$ & $8/21$\\
        &                & $|y|<0.1$ & 5--10\%  & $18.7\pm0.8$ & $0.175\pm0.003$ & $-0.042\pm0.005$ & $0.145\pm0.005$ & $4/21$\\
        &                &           & 10--20\% & $15.3\pm0.6$ & $0.169\pm0.003$ & $-0.030\pm0.005$ & $0.138\pm0.005$ & $3/21$\\
        &                &           & 20--30\% & $11.4\pm0.3$ & $0.164\pm0.003$ & $-0.029\pm0.004$ & $0.128\pm0.005$ & $3/21$\\
        &                &           & 30--40\% & $10.3\pm0.3$ & $0.157\pm0.003$ & $-0.024\pm0.004$ & $0.122\pm0.005$ & $3/21$\\
        &                &           & 40--50\% & $7.9\pm0.2$  & $0.145\pm0.002$ & $-0.021\pm0.004$ & $0.119\pm0.005$ & $3/21$\\
        &                &           & 50--60\% & $7.6\pm0.2$  & $0.138\pm0.002$ & $-0.019\pm0.004$ & $0.117\pm0.005$ & $7/21$\\
        &                &           & 60--70\% & $7.1\pm0.2$  & $0.132\pm0.002$ & $-0.008\pm0.004$ & $0.115\pm0.005$ & $5/21$\\
        &                &           & 70--80\% & $6.1\pm0.1$  & $0.121\pm0.002$ & $-0.003\pm0.004$ & $0.106\pm0.005$ & $9/21$\\
\hline
Figure 6(e) & $\overline{p}$ & $39$  & 0--5\%   & $13.1\pm0.4$ & $0.135\pm0.002$ & $-0.002\pm0.004$ & $0.387\pm0.008$ & $7/18$\\
        &                & $|y|<0.1$ & 5--10\%  & $12.7\pm0.3$ & $0.130\pm0.002$ & $0.003\pm0.004$  & $0.378\pm0.008$ & $5/18$\\
        &                &           & 10--20\% & $11.9\pm0.3$ & $0.128\pm0.002$ & $0.004\pm0.004$  & $0.370\pm0.008$ & $4/18$\\
        &                &           & 20--30\% & $11.3\pm0.3$ & $0.124\pm0.002$ & $0.005\pm0.004$  & $0.340\pm0.008$ & $5/18$\\
        &                &           & 30--40\% & $10.5\pm0.3$ & $0.117\pm0.002$ & $0.007\pm0.004$  & $0.334\pm0.008$ & $3/18$\\
        &                &           & 40--50\% & $9.0\pm0.2$  & $0.111\pm0.002$ & $0.008\pm0.004$  & $0.322\pm0.008$ & $3/18$\\
        &                &           & 50--60\% & $8.7\pm0.2$  & $0.101\pm0.002$ & $0.010\pm0.004$  & $0.316\pm0.008$ & $5/18$\\
        &                &           & 60--70\% & $8.1\pm0.2$  & $0.091\pm0.002$ & $0.012\pm0.004$  & $0.314\pm0.008$ & $7/18$\\
        &                &           & 70--80\% & $8.0\pm0.2$  & $0.088\pm0.001$ & $0.015\pm0.004$  & $0.290\pm0.007$ & $7/18$\\
\hline
Figure 6(f) & $p$        & $39$      & 0--5\%   & $13.9\pm0.4$ & $0.139\pm0.002$ & $-0.007\pm0.004$ & $0.376\pm0.008$ & $3/17$\\
        &                & $|y|<0.1$ & 5--10\%  & $13.1\pm0.4$ & $0.135\pm0.002$ & $-0.004\pm0.004$ & $0.365\pm0.008$ & $5/17$\\
        &                &           & 10--20\% & $11.8\pm0.3$ & $0.133\pm0.002$ & $0.001\pm0.004$  & $0.350\pm0.008$ & $3/17$\\
        &                &           & 20--30\% & $10.3\pm0.3$ & $0.127\pm0.002$ & $0.003\pm0.004$  & $0.341\pm0.008$ & $4/17$\\
        &                &           & 30--40\% & $9.9\pm0.2$  & $0.121\pm0.002$ & $0.005\pm0.004$  & $0.332\pm0.008$ & $8/17$\\
        &                &           & 40--50\% & $9.5\pm0.2$  & $0.120\pm0.002$ & $0.007\pm0.004$  & $0.310\pm0.008$ & $6/17$\\
        &                &           & 50--60\% & $8.9\pm0.2$  & $0.111\pm0.002$ & $0.009\pm0.004$  & $0.303\pm0.008$ & $8/17$\\
        &                &           & 60--70\% & $8.5\pm0.2$  & $0.104\pm0.002$ & $0.011\pm0.004$  & $0.288\pm0.007$ & $9/17$\\
        &                &           & 70--80\% & $9.1\pm0.2$  & $0.100\pm0.002$ & $0.013\pm0.004$  & $0.276\pm0.007$ & $20/17$\\
\hline
\end{tabular}%
\end{center}}
\end{table*}

\begin{figure*}
\begin{center}
\includegraphics[width=14.0cm]{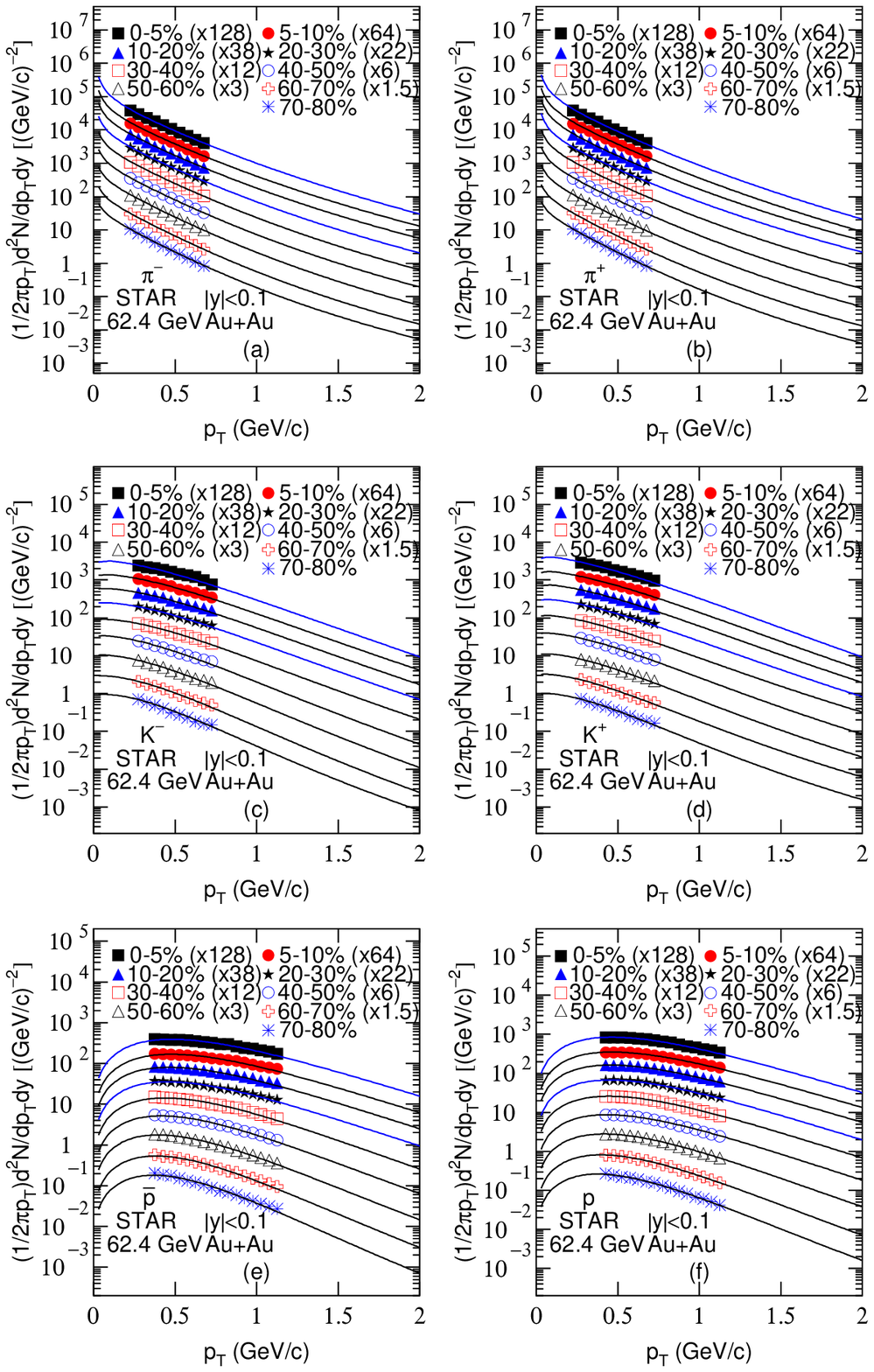}
\end{center}
\justifying\noindent {Figure 7. Same as Figure 1, but showing the
results for 62.4 GeV Au-Au collisions. The symbols represent the
STAR data~\cite{35}.}
\end{figure*}

\begin{table*} \vspace{-0.5cm} \justifying\noindent {\small Table 7.
Values of $n$, $T_{0}$, $a_{0}$, $\langle\beta_{t}\rangle$,
$\chi^{2}$, and ndof corresponding to the curves in Figure 7.
\vspace{-0.5cm}

\begin{center}
\newcommand{\tabincell}[2]{\begin{tabular}{@{}#1@{}}#2\end{tabular}}
\begin{tabular} {cccccccccccc}\\ \hline\hline
Figure & Particle &  $\sqrt{s_{NN}}$ (GeV) & Selection & $n$ & $T_0$ (GeV) & $a_0$ & $\langle\beta_{t}\rangle$ ($c$) & $\chi^2$/ndof \\
\hline
Figure 7(a) & $\pi^{-}$  & $62.4$    & 0--5\%   & $8.9\pm0.2$  & $0.183\pm0.003$ & $-0.419\pm0.005$ & $0.125\pm0.003$ & $82/5$\\
        &                & $|y|<0.1$ & 5--10\%  & $8.3\pm0.2$  & $0.175\pm0.003$ & $-0.416\pm0.005$ & $0.123\pm0.003$ & $83/5$\\
        &                &           & 10--20\% & $7.8\pm0.2$  & $0.171\pm0.003$ & $-0.414\pm0.005$ & $0.122\pm0.003$ & $73/5$\\
        &                &           & 20--30\% & $7.5\pm0.2$  & $0.168\pm0.003$ & $-0.412\pm0.005$ & $0.121\pm0.003$ & $69/5$\\
        &                &           & 30--40\% & $6.7\pm0.1$  & $0.162\pm0.003$ & $-0.410\pm0.005$ & $0.119\pm0.003$ & $62/5$\\
        &                &           & 40--50\% & $6.4\pm0.1$  & $0.159\pm0.003$ & $-0.408\pm0.005$ & $0.118\pm0.003$ & $29/5$\\
        &                &           & 50--60\% & $6.0\pm0.1$  & $0.154\pm0.003$ & $-0.404\pm0.005$ & $0.115\pm0.003$ & $29/5$\\
        &                &           & 60--70\% & $5.7\pm0.1$  & $0.146\pm0.002$ & $-0.394\pm0.005$ & $0.111\pm0.003$ & $20/5$\\
        &                &           & 70--80\% & $5.5\pm0.1$  & $0.139\pm0.002$ & $-0.388\pm0.005$ & $0.106\pm0.003$ & $4/5$\\
\hline
Figure 7(b) & $\pi^{+}$  & $62.4$    & 0--5\%   & $8.2\pm0.2$  & $0.195\pm0.003$ & $-0.426\pm0.005$ & $0.130\pm0.003$ & $65/5$\\
        &                & $|y|<0.1$ & 5--10\%  & $7.5\pm0.2$  & $0.188\pm0.003$ & $-0.424\pm0.005$ & $0.126\pm0.003$ & $103/5$\\
        &                &           & 10--20\% & $6.6\pm0.1$  & $0.178\pm0.003$ & $-0.422\pm0.005$ & $0.125\pm0.003$ & $71/5$\\
        &                &           & 20--30\% & $6.3\pm0.1$  & $0.172\pm0.003$ & $-0.420\pm0.005$ & $0.124\pm0.003$ & $84/5$\\
        &                &           & 30--40\% & $6.1\pm0.1$  & $0.168\pm0.003$ & $-0.418\pm0.005$ & $0.123\pm0.003$ & $52/5$\\
        &                &           & 40--50\% & $6.0\pm0.1$  & $0.161\pm0.003$ & $-0.414\pm0.005$ & $0.121\pm0.003$ & $31/5$\\
        &                &           & 50--60\% & $5.8\pm0.1$  & $0.154\pm0.003$ & $-0.411\pm0.005$ & $0.118\pm0.003$ & $29/5$\\
        &                &           & 60--70\% & $5.6\pm0.1$  & $0.147\pm0.002$ & $-0.407\pm0.005$ & $0.115\pm0.003$ & $20/5$\\
        &                &           & 70--80\% & $5.6\pm0.1$  & $0.140\pm0.002$ & $-0.405\pm0.005$ & $0.113\pm0.003$ & $7/5$\\
\hline
Figure 7(c) & $K^{-}$    & $62.4$    & 0--5\%   & $15.8\pm0.6$ & $0.178\pm0.003$ & $-0.038\pm0.005$ & $0.193\pm0.005$ & $21/5$\\
        &                & $|y|<0.1$ & 5--10\%  & $13.1\pm0.4$ & $0.173\pm0.003$ & $-0.026\pm0.004$ & $0.184\pm0.005$ & $4/5$\\
        &                &           & 10--20\% & $12.3\pm0.3$ & $0.170\pm0.003$ & $-0.018\pm0.004$ & $0.175\pm0.005$ & $8/5$\\
        &                &           & 20--30\% & $11.3\pm0.3$ & $0.166\pm0.003$ & $-0.009\pm0.004$ & $0.166\pm0.005$ & $4/5$\\
        &                &           & 30--40\% & $10.2\pm0.3$ & $0.157\pm0.003$ & $-0.007\pm0.004$ & $0.159\pm0.005$ & $5/5$\\
        &                &           & 40--50\% & $9.7\pm0.2$  & $0.144\pm0.002$ & $-0.022\pm0.004$ & $0.152\pm0.005$ & $12/5$\\
        &                &           & 50--60\% & $9.4\pm0.2$  & $0.136\pm0.002$ & $-0.015\pm0.004$ & $0.143\pm0.005$ & $7/5$\\
        &                &           & 60--70\% & $8.5\pm0.2$  & $0.127\pm0.002$ & $-0.008\pm0.004$ & $0.129\pm0.005$ & $7/5$\\
        &                &           & 70--80\% & $7.1\pm0.2$  & $0.119\pm0.002$ & $0.004\pm0.004$  & $0.119\pm0.005$ & $9/5$\\
\hline
Figure 7(d) & $K^{+}$    & $62.4$    & 0--5\%   & $18.1\pm0.8$ & $0.180\pm0.003$ & $-0.052\pm0.005$ & $0.171\pm0.005$ & $11/5$\\
        &                & $|y|<0.1$ & 5--10\%  & $14.2\pm0.5$ & $0.177\pm0.003$ & $-0.039\pm0.005$ & $0.163\pm0.005$ & $9/5$\\
        &                &           & 10--20\% & $11.1\pm0.3$ & $0.171\pm0.003$ & $-0.030\pm0.005$ & $0.157\pm0.005$ & $8/5$\\
        &                &           & 20--30\% & $9.2\pm0.2$  & $0.166\pm0.003$ & $-0.027\pm0.004$ & $0.149\pm0.005$ & $5/5$\\
        &                &           & 30--40\% & $7.4\pm0.2$  & $0.158\pm0.003$ & $-0.022\pm0.004$ & $0.142\pm0.005$ & $2/5$\\
        &                &           & 40--50\% & $7.3\pm0.2$  & $0.148\pm0.002$ & $-0.017\pm0.004$ & $0.137\pm0.005$ & $4/5$\\
        &                &           & 50--60\% & $6.7\pm0.1$  & $0.140\pm0.002$ & $-0.016\pm0.004$ & $0.132\pm0.005$ & $9/5$\\
        &                &           & 60--70\% & $5.9\pm0.1$  & $0.135\pm0.002$ & $-0.016\pm0.004$ & $0.129\pm0.005$ & $3/5$\\
        &                &           & 70--80\% & $5.4\pm0.1$  & $0.123\pm0.002$ & $-0.012\pm0.004$ & $0.125\pm0.005$ & $8/5$\\
\hline
Figure 7(e) & $\overline{p}$ & $62.4$& 0--5\%   & $12.6\pm0.3$ & $0.142\pm0.002$ & $-0.004\pm0.004$ & $0.428\pm0.008$ & $94/11$\\
        &                & $|y|<0.1$ & 5--10\%  & $12.3\pm0.3$ & $0.138\pm0.002$ & $0.001\pm0.004$  & $0.424\pm0.008$ & $105/11$\\
        &                &           & 10--20\% & $11.4\pm0.3$ & $0.133\pm0.002$ & $0.003\pm0.004$  & $0.410\pm0.008$ & $81/11$\\
        &                &           & 20--30\% & $10.0\pm0.3$ & $0.129\pm0.002$ & $0.003\pm0.004$  & $0.392\pm0.008$ & $114/11$\\
        &                &           & 30--40\% & $10.7\pm0.3$ & $0.121\pm0.002$ & $0.007\pm0.004$  & $0.375\pm0.008$ & $96/11$\\
        &                &           & 40--50\% & $9.2\pm0.2$  & $0.112\pm0.002$ & $0.009\pm0.004$  & $0.349\pm0.008$ & $76/11$\\
        &                &           & 50--60\% & $8.9\pm0.2$  & $0.106\pm0.002$ & $0.011\pm0.004$  & $0.320\pm0.008$ & $50/11$\\
        &                &           & 60--70\% & $8.1\pm0.2$  & $0.096\pm0.002$ & $0.014\pm0.004$  & $0.317\pm0.008$ & $75/11$\\
        &                &           & 70--80\% & $7.8\pm0.2$  & $0.090\pm0.002$ & $0.015\pm0.004$  & $0.303\pm0.008$ & $99/11$\\
\hline
Figure 7(f) & $p$        & $62.4$    & 0--5\%   & $12.7\pm0.3$ & $0.148\pm0.002$ & $-0.009\pm0.004$ & $0.395\pm0.008$ & $29/10$\\
        &                & $|y|<0.1$ & 5--10\%  & $12.5\pm0.3$ & $0.148\pm0.002$ & $-0.005\pm0.004$ & $0.390\pm0.008$ & $14/10$\\
        &                &           & 10--20\% & $10.8\pm0.3$ & $0.142\pm0.002$ & $-0.001\pm0.004$ & $0.381\pm0.008$ & $20/10$\\
        &                &           & 20--30\% & $11.0\pm0.3$ & $0.138\pm0.002$ & $0.004\pm0.004$  & $0.372\pm0.008$ & $18/10$\\
        &                &           & 30--40\% & $11.1\pm0.3$ & $0.136\pm0.002$ & $0.005\pm0.004$  & $0.363\pm0.008$ & $10/10$\\
        &                &           & 40--50\% & $10.2\pm0.3$ & $0.129\pm0.002$ & $0.008\pm0.004$  & $0.350\pm0.008$ & $12/10$\\
        &                &           & 50--60\% & $9.9\pm0.2$  & $0.117\pm0.002$ & $0.009\pm0.004$  & $0.344\pm0.008$ & $9/10$\\
        &                &           & 60--70\% & $9.2\pm0.2$  & $0.108\pm0.002$ & $0.010\pm0.004$  & $0.321\pm0.008$ & $9/10$\\
        &                &           & 70--80\% & $8.2\pm0.2$  & $0.101\pm0.002$ & $0.011\pm0.004$  & $0.297\pm0.007$ & $25/10$\\
\hline
\end{tabular}%
\end{center}}
\end{table*}

\begin{figure*}
\begin{center}
\includegraphics[width=14.0cm]{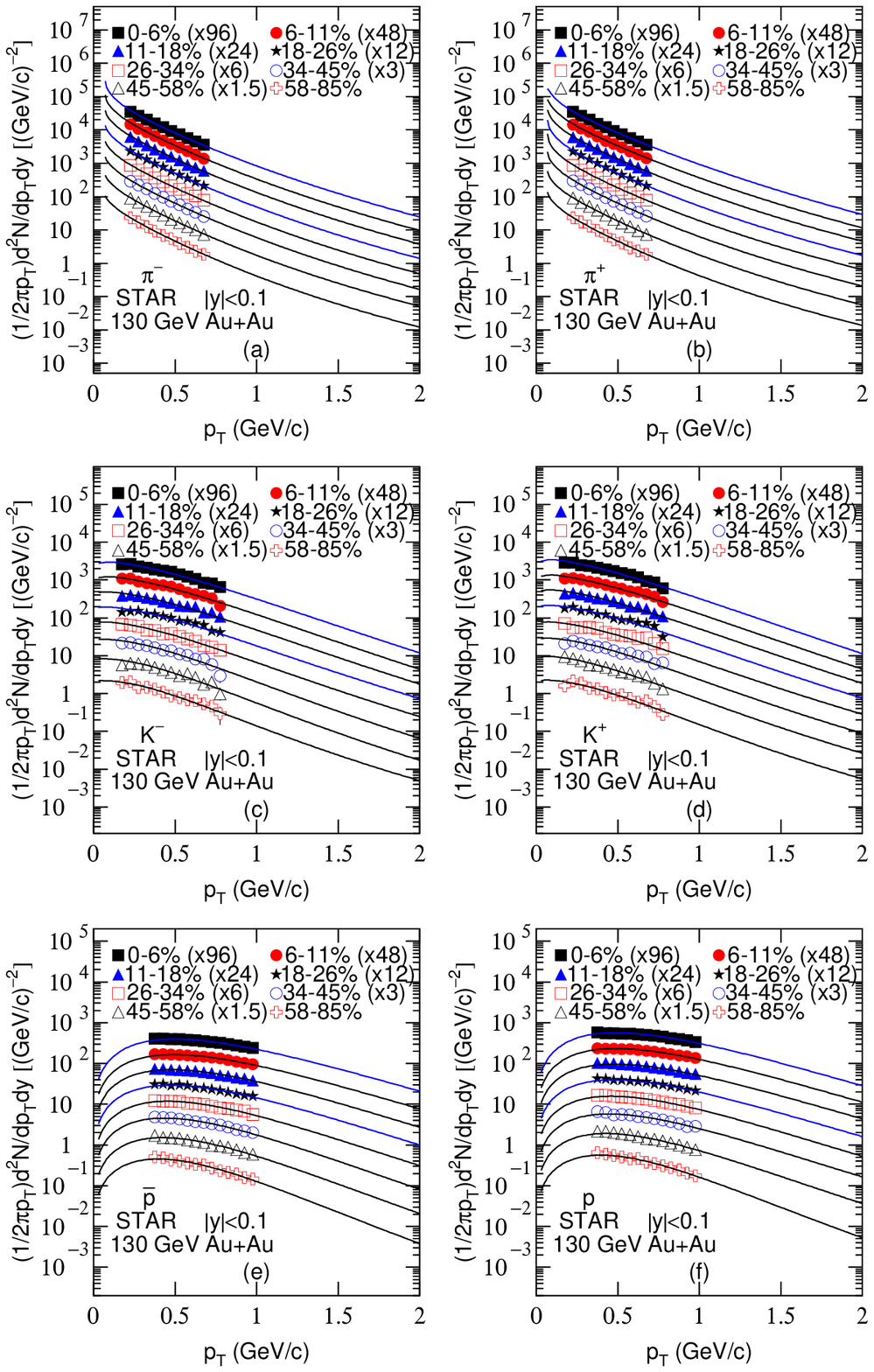}
\end{center}
\justifying\noindent {Figure 8. Same as Figure 1, but showing the
results for 130 GeV Au-Au collisions with another set of
centrality intervals. The symbols represent the STAR
data~\cite{35}.}
\end{figure*}

\begin{table*} \vspace{0.5cm} \justifying\noindent {\small Table 8.
Values of $n$, $T_{0}$, $a_{0}$, $\langle\beta_{t}\rangle$,
$\chi^{2}$, and ndof corresponding to the curves in Figure 8.
\vspace{-0.5cm}

\begin{center}
\newcommand{\tabincell}[2]{\begin{tabular}{@{}#1@{}}#2\end{tabular}}
\begin{tabular} {cccccccccccc}\\ \hline\hline
Figure & Particle &  $\sqrt{s_{NN}}$ (GeV) & Selection & $n$ & $T_0$ (GeV) & $a_0$ & $\langle\beta_{t}\rangle$ ($c$) & $\chi^2$/ndof \\
\hline
Figure 8(a) & $\pi^{-}$  & $130$     & 0--6\%   & $7.2\pm0.2$  & $0.185\pm0.003$ & $-0.435\pm0.005$ & $0.131\pm0.003$ & $54/5$\\
        &                & $|y|<0.1$ & 6--11\%  & $6.8\pm0.1$  & $0.177\pm0.003$ & $-0.432\pm0.005$ & $0.124\pm0.003$ & $46/5$\\
        &                &           & 11--18\% & $6.6\pm0.1$  & $0.173\pm0.003$ & $-0.431\pm0.005$ & $0.122\pm0.003$ & $34/5$\\
        &                &           & 18--26\% & $6.4\pm0.1$  & $0.170\pm0.003$ & $-0.430\pm0.005$ & $0.121\pm0.003$ & $17/5$\\
        &                &           & 26--34\% & $6.2\pm0.1$  & $0.165\pm0.003$ & $-0.427\pm0.005$ & $0.119\pm0.003$ & $12/5$\\
        &                &           & 34--45\% & $6.1\pm0.1$  & $0.163\pm0.003$ & $-0.426\pm0.005$ & $0.118\pm0.003$ & $14/5$\\
        &                &           & 45--58\% & $5.5\pm0.1$  & $0.157\pm0.003$ & $-0.424\pm0.005$ & $0.115\pm0.003$ & $9/5$\\
        &                &           & 58--85\% & $5.3\pm0.1$  & $0.150\pm0.003$ & $-0.414\pm0.005$ & $0.113\pm0.003$ & $8/5$\\
\hline
Figure 8(b) & $\pi^{+}$  & $130$     & 0--6\%   & $8.0\pm0.2$  & $0.198\pm0.003$ & $-0.429\pm0.005$ & $0.137\pm0.003$ & $14/5$\\
        &                & $|y|<0.1$ & 6--11\%  & $7.3\pm0.2$  & $0.191\pm0.003$ & $-0.427\pm0.005$ & $0.134\pm0.003$ & $13/5$\\
        &                &           & 11--18\% & $6.4\pm0.1$  & $0.181\pm0.003$ & $-0.424\pm0.005$ & $0.132\pm0.003$ & $14/5$\\
        &                &           & 18--26\% & $6.1\pm0.1$  & $0.177\pm0.003$ & $-0.417\pm0.005$ & $0.130\pm0.003$ & $18/5$\\
        &                &           & 26--34\% & $5.7\pm0.1$  & $0.174\pm0.003$ & $-0.416\pm0.005$ & $0.126\pm0.003$ & $13/5$\\
        &                &           & 34--45\% & $5.4\pm0.1$  & $0.164\pm0.003$ & $-0.413\pm0.005$ & $0.125\pm0.003$ & $9/5$\\
        &                &           & 45--58\% & $5.4\pm0.1$  & $0.159\pm0.003$ & $-0.411\pm0.005$ & $0.124\pm0.003$ & $15/5$\\
        &                &           & 58--85\% & $5.0\pm0.1$  & $0.150\pm0.003$ & $-0.410\pm0.005$ & $0.123\pm0.003$ & $18/5$\\
\hline
Figure 8(c) & $K^{-}$    & $130$     & 0--6\%   & $14.3\pm0.5$ & $0.183\pm0.003$ & $-0.038\pm0.005$ & $0.209\pm0.005$ & $7/8$\\
        &                & $|y|<0.1$ & 6--11\%  & $12.9\pm0.3$ & $0.177\pm0.003$ & $-0.027\pm0.004$ & $0.196\pm0.005$ & $9/8$\\
        &                &           & 11--18\% & $10.8\pm0.3$ & $0.174\pm0.003$ & $-0.019\pm0.004$ & $0.181\pm0.005$ & $5/8$\\
        &                &           & 18--26\% & $8.3\pm0.2$  & $0.168\pm0.003$ & $-0.016\pm0.004$ & $0.167\pm0.005$ & $12/8$\\
        &                &           & 26--34\% & $8.0\pm0.2$  & $0.161\pm0.003$ & $-0.014\pm0.004$ & $0.159\pm0.005$ & $7/8$\\
        &                &           & 34--45\% & $7.6\pm0.2$  & $0.148\pm0.002$ & $-0.007\pm0.004$ & $0.154\pm0.005$ & $10/8$\\
        &                &           & 45--58\% & $7.1\pm0.2$  & $0.141\pm0.002$ & $-0.003\pm0.004$ & $0.144\pm0.005$ & $14/8$\\
        &                &           & 58--85\% & $5.6\pm0.1$  & $0.132\pm0.002$ & $0.001\pm0.004$  & $0.137\pm0.005$ & $7/8$\\
\hline
Figure 8(d) & $K^{+}$    & $130$     & 0--6\%   & $15.9\pm0.6$ & $0.186\pm0.003$ & $-0.054\pm0.005$ & $0.194\pm0.005$ & $10/8$\\
        &                & $|y|<0.1$ & 6--11\%  & $14.1\pm0.5$ & $0.181\pm0.003$ & $-0.041\pm0.005$ & $0.188\pm0.005$ & $10/8$\\
        &                &           & 11--18\% & $11.2\pm0.3$ & $0.175\pm0.003$ & $-0.032\pm0.005$ & $0.182\pm0.005$ & $7/8$\\
        &                &           & 18--26\% & $9.3\pm0.2$  & $0.170\pm0.003$ & $-0.022\pm0.004$ & $0.171\pm0.005$ & $12/8$\\
        &                &           & 26--34\% & $7.6\pm0.2$  & $0.163\pm0.003$ & $-0.017\pm0.004$ & $0.167\pm0.005$ & $17/8$\\
        &                &           & 34--45\% & $6.9\pm0.1$  & $0.152\pm0.003$ & $-0.013\pm0.004$ & $0.162\pm0.005$ & $15/8$\\
        &                &           & 45--58\% & $6.3\pm0.1$  & $0.144\pm0.002$ & $-0.011\pm0.004$ & $0.159\pm0.005$ & $7/8$\\
        &                &           & 58--85\% & $5.9\pm0.1$  & $0.138\pm0.002$ & $-0.009\pm0.004$ & $0.147\pm0.005$ & $15/8$\\
\hline
Figure 8(e) & $\overline{p}$ & $130$ & 0--6\%   & $12.1\pm0.3$ & $0.146\pm0.002$ & $-0.006\pm0.004$ & $0.457\pm0.008$ & $81/8$\\
        &                & $|y|<0.1$ & 6--11\%  & $11.9\pm0.3$ & $0.143\pm0.002$ & $-0.002\pm0.004$ & $0.444\pm0.008$ & $33/8$\\
        &                &           & 11--18\% & $11.3\pm0.3$ & $0.137\pm0.002$ & $0.002\pm0.004$  & $0.429\pm0.008$ & $12/8$\\
        &                &           & 18--26\% & $10.7\pm0.3$ & $0.133\pm0.002$ & $0.003\pm0.004$  & $0.421\pm0.008$ & $32/8$\\
        &                &           & 26--34\% & $9.3\pm0.2$  & $0.125\pm0.002$ & $0.007\pm0.004$  & $0.398\pm0.008$ & $25/8$\\
        &                &           & 34--45\% & $9.0\pm0.2$  & $0.117\pm0.002$ & $0.008\pm0.004$  & $0.375\pm0.008$ & $26/8$\\
        &                &           & 45--58\% & $8.7\pm0.2$  & $0.110\pm0.002$ & $0.011\pm0.004$  & $0.367\pm0.008$ & $25/8$\\
        &                &           & 58--85\% & $7.9\pm0.2$  & $0.102\pm0.002$ & $0.012\pm0.004$  & $0.328\pm0.008$ & $22/8$\\
\hline
Figure 8(f) & $p$        & $130$     & 0--6\%   & $12.3\pm0.4$ & $0.149\pm0.002$ & $-0.009\pm0.004$ & $0.439\pm0.008$ & $66/8$\\
        &                & $|y|<0.1$ & 6--11\%  & $12.0\pm0.4$ & $0.148\pm0.002$ & $-0.006\pm0.004$ & $0.432\pm0.008$ & $41/8$\\
        &                &           & 11--18\% & $10.6\pm0.3$ & $0.144\pm0.002$ & $-0.002\pm0.004$ & $0.425\pm0.008$ & $13/8$\\
        &                &           & 18--26\% & $9.2\pm0.2$  & $0.139\pm0.002$ & $0.001\pm0.004$  & $0.411\pm0.008$ & $6/8$\\
        &                &           & 26--34\% & $8.9\pm0.2$  & $0.133\pm0.002$ & $0.004\pm0.004$  & $0.398\pm0.008$ & $16/8$\\
        &                &           & 34--45\% & $8.1\pm0.2$  & $0.129\pm0.002$ & $0.007\pm0.004$  & $0.375\pm0.008$ & $24/8$\\
        &                &           & 45--58\% & $7.7\pm0.2$  & $0.117\pm0.002$ & $0.008\pm0.004$  & $0.363\pm0.008$ & $10/8$\\
        &                &           & 58--85\% & $7.4\pm0.2$  & $0.102\pm0.002$ & $0.010\pm0.004$  & $0.333\pm0.008$ & $11/8$\\
\hline
\end{tabular}%
\end{center}}
\end{table*}

\begin{figure*}
\begin{center}
\includegraphics[width=14.0cm]{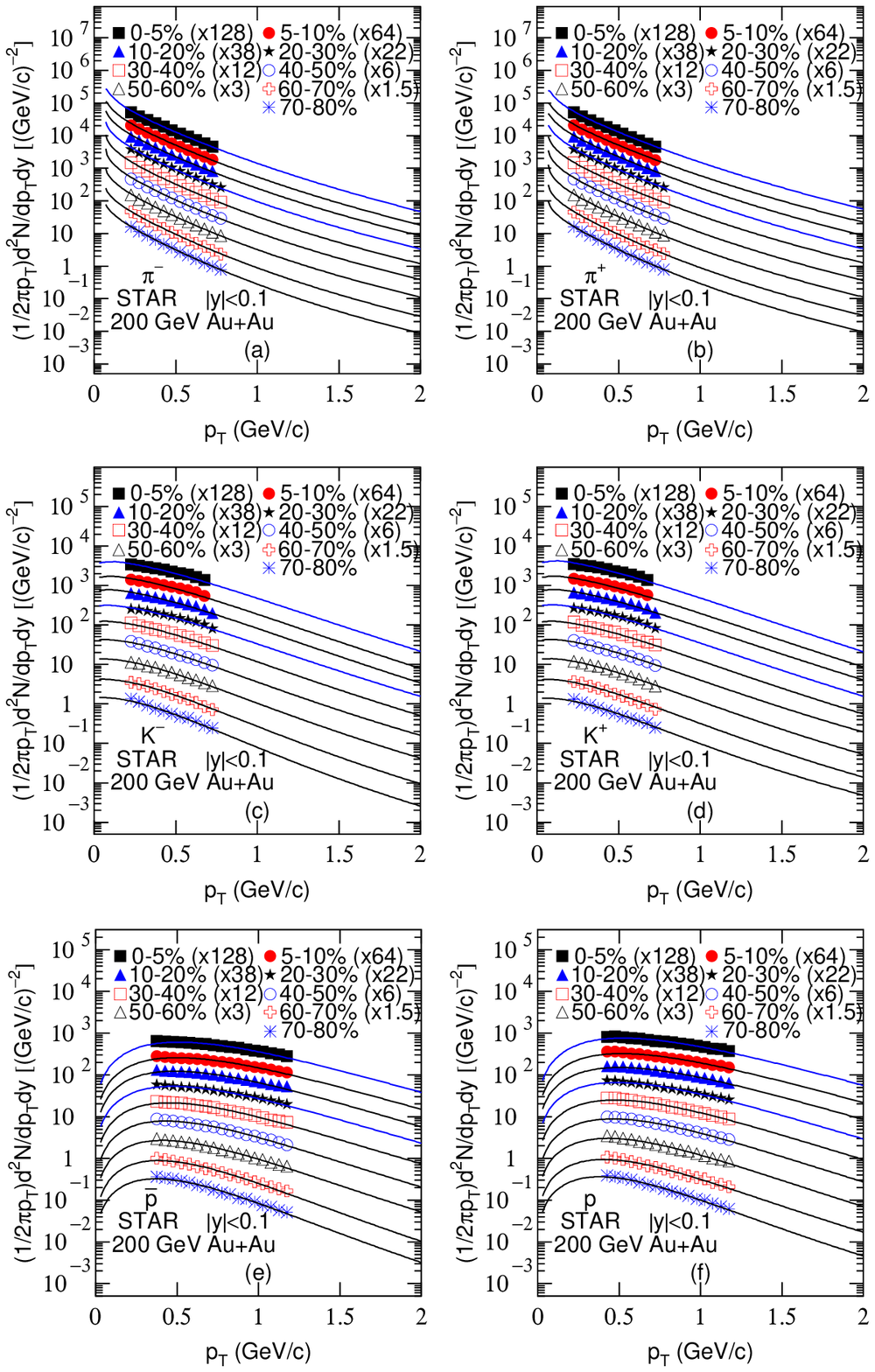}
\end{center}
\justifying\noindent {Figure 9. Same as Figure 1, but showing the
results for 200 GeV Au-Au collisions. The symbols represent the
STAR data~\cite{35}.}
\end{figure*}

\begin{table*} \vspace{-0.5cm} \justifying\noindent {\small Table 9.
Values of $n$, $T_{0}$, $a_{0}$, $\langle\beta_{t}\rangle$,
$\chi^{2}$, and ndof corresponding to the curves in Figure 9.
\vspace{-0.5cm}

\begin{center}
\newcommand{\tabincell}[2]{\begin{tabular}{@{}#1@{}}#2\end{tabular}}
\begin{tabular} {cccccccccccc}\\ \hline\hline
Figure & Particle &  $\sqrt{s_{NN}}$ (GeV) & Selection & $n$ & $T_0$ (GeV) & $a_0$ & $\langle\beta_{t}\rangle$ ($c$) & $\chi^2$/ndof \\
\hline
Figure 9(a) & $\pi^{-}$  & $200$     & 0--5\%   & $6.7\pm0.1$ & $0.189\pm0.003$ & $-0.439\pm0.005$ & $0.139\pm0.003$ & $58/6$\\
        &                & $|y|<0.1$ & 5--10\%  & $6.1\pm0.1$ & $0.180\pm0.003$ & $-0.437\pm0.005$ & $0.138\pm0.003$ & $50/6$\\
        &                &           & 10--20\% & $5.7\pm0.1$ & $0.178\pm0.003$ & $-0.434\pm0.005$ & $0.135\pm0.003$ & $77/6$\\
        &                &           & 20--30\% & $5.5\pm0.1$ & $0.174\pm0.003$ & $-0.432\pm0.005$ & $0.133\pm0.003$ & $64/7$\\
        &                &           & 30--40\% & $5.2\pm0.1$ & $0.168\pm0.003$ & $-0.430\pm0.005$ & $0.132\pm0.003$ & $54/7$\\
        &                &           & 40--50\% & $5.1\pm0.1$ & $0.167\pm0.003$ & $-0.426\pm0.005$ & $0.125\pm0.003$ & $46/7$\\
        &                &           & 50--60\% & $5.0\pm0.1$ & $0.160\pm0.003$ & $-0.423\pm0.005$ & $0.123\pm0.003$ & $33/7$\\
        &                &           & 60--70\% & $5.0\pm0.1$ & $0.153\pm0.003$ & $-0.420\pm0.005$ & $0.120\pm0.003$ & $13/7$\\
        &                &           & 70--80\% & $4.9\pm0.1$ & $0.145\pm0.002$ & $-0.419\pm0.005$ & $0.116\pm0.003$ & $5/7$\\
\hline
Figure 9(b) & $\pi^{+}$  & $200$     & 0--5\%   & $7.0\pm0.1$ & $0.201\pm0.003$ & $-0.432\pm0.005$ & $0.141\pm0.003$ & $29/6$\\
        &                & $|y|<0.1$ & 5--10\%  & $6.4\pm0.1$ & $0.194\pm0.003$ & $-0.426\pm0.005$ & $0.139\pm0.003$ & $24/6$\\
        &                &           & 10--20\% & $6.3\pm0.1$ & $0.184\pm0.003$ & $-0.421\pm0.005$ & $0.137\pm0.003$ & $42/6$\\
        &                &           & 20--30\% & $6.1\pm0.1$ & $0.180\pm0.003$ & $-0.418\pm0.005$ & $0.135\pm0.003$ & $25/7$\\
        &                &           & 30--40\% & $5.7\pm0.1$ & $0.177\pm0.003$ & $-0.416\pm0.005$ & $0.132\pm0.003$ & $16/7$\\
        &                &           & 40--50\% & $5.5\pm0.1$ & $0.167\pm0.003$ & $-0.412\pm0.005$ & $0.131\pm0.003$ & $31/7$\\
        &                &           & 50--60\% & $5.2\pm0.1$ & $0.162\pm0.003$ & $-0.406\pm0.005$ & $0.130\pm0.003$ & $26/7$\\
        &                &           & 60--70\% & $4.9\pm0.1$ & $0.153\pm0.003$ & $-0.403\pm0.005$ & $0.125\pm0.003$ & $11/7$\\
        &                &           & 70--80\% & $4.8\pm0.1$ & $0.146\pm0.002$ & $-0.401\pm0.005$ & $0.124\pm0.003$ & $2/7$\\
\hline
Figure 9(c) & $K^{-}$    & $200$     & 0--5\%   & $12.4\pm0.3$ & $0.187\pm0.003$ & $-0.040\pm0.005$ & $0.232\pm0.005$ & $9/5$\\
        &                & $|y|<0.1$ & 5--10\%  & $11.3\pm0.3$ & $0.181\pm0.003$ & $-0.033\pm0.005$ & $0.212\pm0.005$ & $8/5$\\
        &                &           & 10--20\% & $10.4\pm0.3$ & $0.178\pm0.003$ & $-0.026\pm0.004$ & $0.195\pm0.005$ & $8/6$\\
        &                &           & 20--30\% & $7.9\pm0.2$  & $0.174\pm0.003$ & $-0.018\pm0.004$ & $0.183\pm0.005$ & $8/6$\\
        &                &           & 30--40\% & $7.4\pm0.2$  & $0.165\pm0.003$ & $-0.015\pm0.004$ & $0.173\pm0.005$ & $5/6$\\
        &                &           & 40--50\% & $6.0\pm0.1$  & $0.152\pm0.003$ & $-0.010\pm0.004$ & $0.162\pm0.005$ & $5/6$\\
        &                &           & 50--60\% & $5.9\pm0.1$  & $0.145\pm0.002$ & $-0.007\pm0.004$ & $0.150\pm0.005$ & $4/6$\\
        &                &           & 60--70\% & $5.9\pm0.1$  & $0.136\pm0.002$ & $-0.004\pm0.004$ & $0.136\pm0.005$ & $4/6$\\
        &                &           & 70--80\% & $5.6\pm0.1$  & $0.127\pm0.002$ & $-0.003\pm0.004$ & $0.128\pm0.005$ & $12/6$\\
\hline
Figure 9(d) & $K^{+}$    & $200$     & 0--5\%   & $13.9\pm0.3$ & $0.188\pm0.003$ & $-0.053\pm0.005$ & $0.239\pm0.005$ & $10/5$\\
        &                &           & 5--10\%  & $11.6\pm0.3$ & $0.183\pm0.003$ & $-0.039\pm0.005$ & $0.227\pm0.005$ & $8/5$\\
        &                &           & 10--20\% & $9.6\pm0.2$  & $0.178\pm0.003$ & $-0.030\pm0.005$ & $0.217\pm0.005$ & $6/6$\\
        &                &           & 20--30\% & $8.7\pm0.2$  & $0.173\pm0.003$ & $-0.026\pm0.004$ & $0.208\pm0.005$ & $5/6$\\
        &                &           & 30--40\% & $7.7\pm0.2$  & $0.166\pm0.003$ & $-0.020\pm0.004$ & $0.197\pm0.005$ & $4/6$\\
        &                &           & 40--50\% & $6.0\pm0.1$  & $0.156\pm0.003$ & $-0.017\pm0.004$ & $0.187\pm0.005$ & $4/6$\\
        &                &           & 50--60\% & $5.8\pm0.1$  & $0.148\pm0.002$ & $-0.014\pm0.004$ & $0.176\pm0.005$ & $3/6$\\
        &                &           & 60--70\% & $5.5\pm0.1$  & $0.143\pm0.002$ & $-0.012\pm0.004$ & $0.162\pm0.005$ & $5/6$\\
        &                &           & 70--80\% & $4.4\pm0.1$  & $0.131\pm0.002$ & $-0.007\pm0.004$ & $0.155\pm0.005$ & $6/6$\\
\hline
Figure 9(e) & $\overline{p}$ & $200$ & 0--5\%   & $11.8\pm0.3$ & $0.151\pm0.003$ & $-0.012\pm0.004$ & $0.472\pm0.008$ & $79/12$\\
        &                & $|y|<0.1$ & 5--10\%  & $10.5\pm0.3$ & $0.147\pm0.002$ & $-0.011\pm0.004$ & $0.464\pm0.008$ & $89/12$\\
        &                &           & 10--20\% & $10.3\pm0.3$ & $0.141\pm0.002$ & $-0.010\pm0.004$ & $0.446\pm0.008$ & $80/12$\\
        &                &           & 20--30\% & $8.4\pm0.2$  & $0.138\pm0.002$ & $-0.010\pm0.004$ & $0.428\pm0.008$ & $89/12$\\
        &                &           & 30--40\% & $6.8\pm0.1$  & $0.129\pm0.002$ & $-0.009\pm0.004$ & $0.399\pm0.008$ & $57/12$\\
        &                &           & 40--50\% & $6.3\pm0.1$  & $0.121\pm0.002$ & $-0.007\pm0.004$ & $0.374\pm0.008$ & $74/12$\\
        &                &           & 50--60\% & $6.2\pm0.1$  & $0.115\pm0.002$ & $-0.005\pm0.004$ & $0.347\pm0.008$ & $44/12$\\
        &                &           & 60--70\% & $5.9\pm0.1$  & $0.104\pm0.002$ & $-0.005\pm0.004$ & $0.317\pm0.008$ & $69/12$\\
        &                &           & 70--80\% & $5.8\pm0.1$  & $0.103\pm0.002$ & $-0.002\pm0.004$ & $0.289\pm0.007$ & $43/12$\\
\hline
Figure 9(f) & $p$        & $200$     & 0--5\%   & $11.7\pm0.3$ & $0.154\pm0.003$ & $-0.011\pm0.004$ & $0.492\pm0.008$ & $30/11$\\
        &                & $|y|<0.1$ & 5--10\%  & $11.2\pm0.3$ & $0.151\pm0.003$ & $-0.008\pm0.004$ & $0.478\pm0.008$ & $35/11$\\
        &                &           & 10--20\% & $10.8\pm0.3$ & $0.145\pm0.002$ & $-0.005\pm0.004$ & $0.466\pm0.008$ & $23/11$\\
        &                &           & 20--30\% & $8.8\pm0.2$  & $0.133\pm0.002$ & $0.002\pm0.004$  & $0.449\pm0.008$ & $31/11$\\
        &                &           & 30--40\% & $8.1\pm0.2$  & $0.127\pm0.002$ & $0.003\pm0.004$  & $0.436\pm0.008$ & $29/11$\\
        &                &           & 40--50\% & $7.5\pm0.2$  & $0.122\pm0.002$ & $0.004\pm0.004$  & $0.417\pm0.008$ & $30/11$\\
        &                &           & 50--60\% & $6.8\pm0.1$  & $0.118\pm0.002$ & $0.006\pm0.004$  & $0.371\pm0.008$ & $44/11$\\
        &                &           & 60--70\% & $5.3\pm0.1$  & $0.107\pm0.002$ & $0.008\pm0.004$  & $0.338\pm0.008$ & $36/11$\\
        &                &           & 70--80\% & $5.0\pm0.1$  & $0.100\pm0.002$ & $0.010\pm0.004$  & $0.300\pm0.007$ & $14/11$\\
\hline
\end{tabular}%
\end{center}}
\end{table*}

\clearpage

\begin{figure*}
\begin{center}
\includegraphics[width=14.0cm]{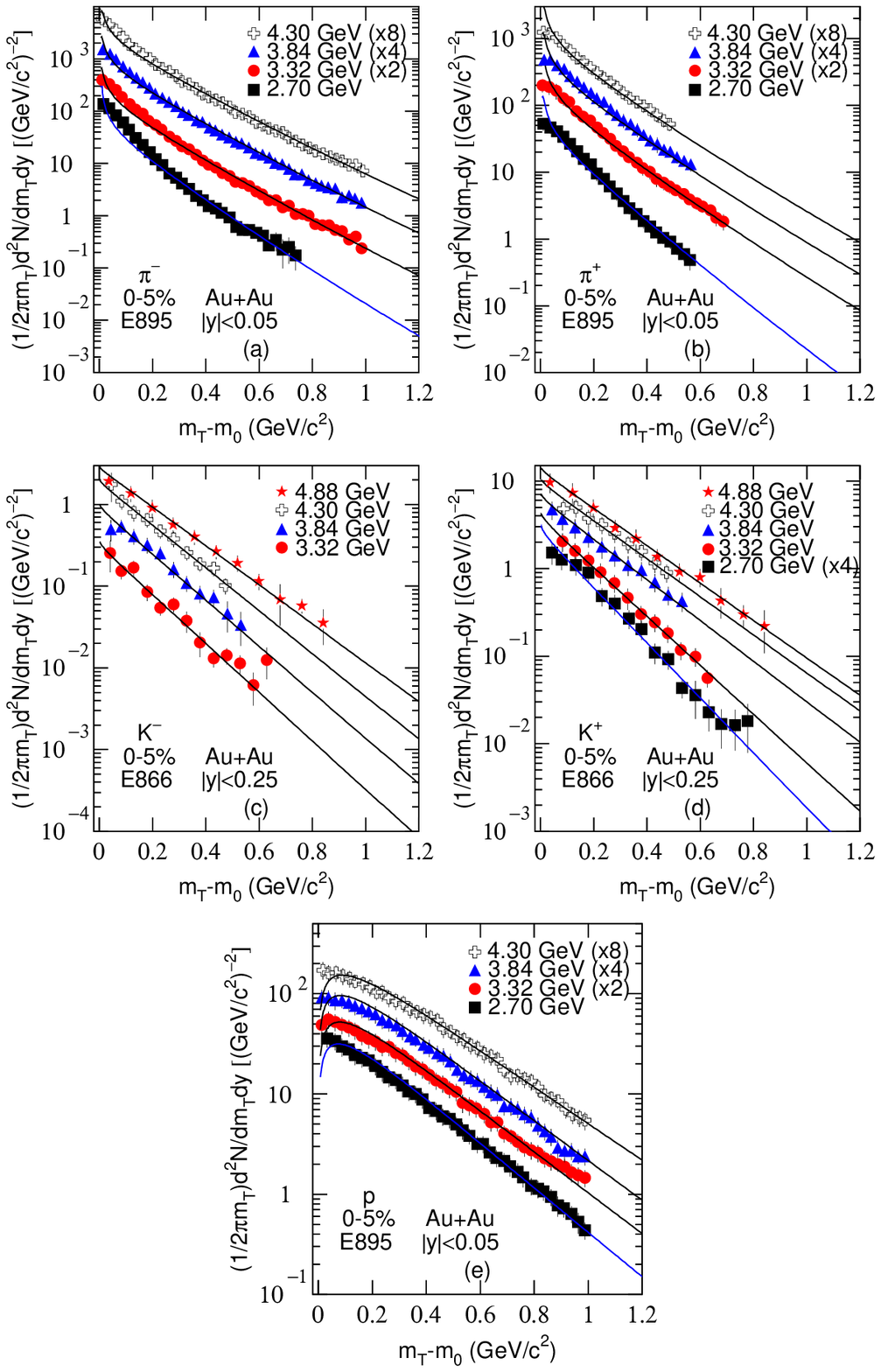}
\end{center}
\justifying\noindent {Figure 10. Transverse mass spectra of
$\pi^{-}$ (a), $\pi^{+}$ (b), $K^{-}$ (c), $K^{+}$ (d), and $p$
(e) produced in 0--5\% Au-Au collisions at mid-rapidity at
$\sqrt{s_{NN}}=2.70$, 3.32, 3.84, 4.30, and 4.88 GeV, where only
panel (d) contains five energies. The symbols in panels (a), (b),
and (e) represent the experimental data measured by the E895
Collaboration~\cite{36}, and those in panels (c) and (d) are from
the E866 Collaboration~\cite{37}, where some sets of data are
re-scaled by different amounts. The data for $\overline{p}$ is not
available in the same or similar experiments. The curves are our
results fitted by Eq. (4) for $\pi^{\mp}$ and $K^{\mp}$ [panels
(a)--(d)] or Eq. (6) for $p$ [panel (e)].}
\end{figure*}

\begin{table*} \vspace{-0.5cm} \justifying\noindent {\small Table 10.
Values of $n$, $T_{0}$, $a_{0}$, $\langle\beta_{t}\rangle$,
$\chi^{2}$, and ndof corresponding to the curves in Figure 10.
\vspace{-0.5cm}

\begin{center}
\newcommand{\tabincell}[2]{\begin{tabular}{@{}#1@{}}#2\end{tabular}}
\begin{tabular} {cccccccccccc}\\ \hline\hline
Figure & Particle &  $\sqrt{s_{NN}}$ (GeV) & Selection & $n$ & $T_0$ (GeV) & $a_0$ & $\langle\beta_{t}\rangle$ ($c$) & $\chi^2$/ndof \\
\hline
Figure 10(a) & $\pi^{-}$ & $2.70 $ & 0--5\%   & $20.6\pm1.0$ & $0.104\pm0.002$ & $-0.455\pm0.005$ & $0.098\pm0.002$ & $35/25$\\
             &           & $3.32 $ &$|y|<0.05$& $19.9\pm0.9$ & $0.127\pm0.002$ & $-0.441\pm0.005$ & $0.105\pm0.003$ & $30/35$\\
             &           & $3.84 $ &          & $19.0\pm0.8$ & $0.135\pm0.002$ & $-0.430\pm0.005$ & $0.107\pm0.003$ & $13/35$\\
             &           & $4.30 $ &          & $18.2\pm0.8$ & $0.141\pm0.002$ & $-0.424\pm0.005$ & $0.110\pm0.003$ & $13/35$\\
\hline
Figure 10(b) & $\pi^{+}$ & $2.70 $ & 0--5\%   & $21.3\pm1.1$ & $0.107\pm0.002$ & $-0.463\pm0.005$ & $0.110\pm0.003$ & $53/18$\\
             &           & $3.32 $ &$|y|<0.05$& $20.5\pm1.0$ & $0.137\pm0.002$ & $-0.458\pm0.005$ & $0.112\pm0.003$ & $56/23$\\
             &           & $3.84 $ &          & $19.8\pm0.9$ & $0.144\pm0.002$ & $-0.443\pm0.005$ & $0.114\pm0.003$ & $54/18$\\
             &           & $4.30 $ &          & $19.0\pm0.8$ & $0.147\pm0.002$ & $-0.438\pm0.005$ & $0.116\pm0.003$ & $61/15$\\
\hline
Figure 10(c) & $K^{-}$   & $3.32 $ & 0--5\%   & $31.7\pm2.1$ & $0.108\pm0.002$ & $-0.044\pm0.005$ & $0.092\pm0.004$ & $9/8$\\
             &           & $3.84 $ &$|y|<0.25$& $30.4\pm2.0$ & $0.114\pm0.002$ & $-0.037\pm0.005$ & $0.095\pm0.004$ & $4/6$\\
             &           & $4.30 $ &          & $29.6\pm1.9$ & $0.122\pm0.002$ & $-0.028\pm0.004$ & $0.099\pm0.004$ & $1/5$\\
             &           & $4.88 $ &          & $28.8\pm1.8$ & $0.134\pm0.002$ & $-0.020\pm0.004$ & $0.103\pm0.005$ & $1/6$\\
\hline
Figure 10(d) & $K^{+}$   & $2.70 $ & 0--5\%   & $33.5\pm2.3$ & $0.102\pm0.002$ & $-0.064\pm0.005$ & $0.103\pm0.005$ & $6/11$\\
             &           & $3.32 $ &$|y|<0.25$& $32.4\pm2.2$ & $0.115\pm0.002$ & $-0.057\pm0.005$ & $0.109\pm0.005$ & $1/7$\\
             &           & $3.84 $ &          & $31.8\pm2.1$ & $0.140\pm0.002$ & $-0.048\pm0.005$ & $0.113\pm0.005$ & $1/6$\\
             &           & $4.30 $ &          & $31.0\pm2.0$ & $0.149\pm0.002$ & $-0.039\pm0.005$ & $0.116\pm0.005$ & $1/4$\\
             &           & $4.88 $ &          & $30.3\pm2.0$ & $0.151\pm0.002$ & $-0.036\pm0.005$ & $0.119\pm0.005$ & $1/6$\\
\hline
Figure 10(e) & $p$       & $2.70 $ & 0--5\%   & $22.0\pm1.1$ & $0.106\pm0.002$ & $-0.016\pm0.004$ & $0.324\pm0.005$ & $8/34$\\
             &           & $3.32 $ &$|y|<0.05$& $21.3\pm1.1$ & $0.114\pm0.002$ & $-0.013\pm0.004$ & $0.330\pm0.005$ & $14/35$\\
             &           & $3.84 $ &          & $20.5\pm1.0$ & $0.116\pm0.002$ & $-0.010\pm0.004$ & $0.334\pm0.005$ & $9/35$\\
             &           & $4.30 $ &          & $19.6\pm0.9$ & $0.122\pm0.002$ & $-0.003\pm0.004$ & $0.342\pm0.005$ & $9/35$\\
\hline
\end{tabular}%
\end{center}}
\end{table*}

\subsection{Tendencies of parameters}

We now analyze the tendencies of parameters on collision energy
$\sqrt{s_{NN}}$ and centrality interval (percentage). In Figure
11, the dependence of parameter $n$ on $\sqrt{s_{NN}}$ and
centrality is shown, where panels (a)--(f) are for the results
from the spectra of $\pi^-$, $\pi^+$, $K^-$, $K^+$,
$\overline{p}$, and $p$, respectively. It is not difficult to find
out that with the increase of $\sqrt{s_{NN}}$, the values of $n$
decrease significantly. Meanwhile, with the decrease of
centrality from central to peripheral collisions, the values
of $n$ also decrease. As we know, $n$ is used to describe
the degree of non-equilibrium of the system. The larger
the value of $n$ is, the closer to equilibrium the system is.
This work shows that the system at lower energy and in central
collisions corresponds to larger $n$ and hence to higher degree of
equilibrium, compared with the system at higher energy and in
peripheral collisions.

As one of the core concepts in thermodynamics and statistical
mechanics, temperature is also the attention of our research.
Generally, we can directly extract the temperature information of
the system from particle spectra. Panels (a)--(f) in Figure 12
show the dependence of thermal freeze-out temperature $T_{0}$ on
$\sqrt{s_{NN}}$ extracted from the spectra of six kinds of charged
hadrons in different centrality intervals. It can be clearly seen
that in the available central collisions, $T_{0}$ increases
rapidly in $\sqrt{s_{NN}}=2.7$--7.7 GeV range. When
$\sqrt{s_{NN}}$ is larger than 7.7 GeV, $T_0$ increases slowly.
The higher the temperature is, the higher the excitation degree
is. As $\sqrt{s_{NN}}$ increases, the excitation degree of the
system also increases. The change trend of $T_0$ with
$\sqrt{s_{NN}}$ can well reflect this change phenomenon of the
excitation degree. At nine collision energies and for six kinds of
particles, $T_{0}$ decreases with the decrease of the centrality
from central to peripheral collisions.

Similar to Figures 11 and 12, the dependence of the dimensionless
correction index $a_{0}$ on $\sqrt{s_{NN}}$ is shown in Figures
13(a)--13(f) for different centrality intervals and charged
hadrons. From Figure 13 one can see that with the increase of
$\sqrt{s_{NN}}$, in the available central collisions, $a_0$
increases quickly and then decreases significantly, and the
boundary energy is 7.7 GeV. The values of $a_0$ from $\pi^{\mp}$
spectra are negative, which is significantly less than those from
$K^{\mp}$ and $\overline{p}$($p$) spectra. From central to
peripheral collisions, $a_0$ increases significantly for different
kinds of particles in the available energy range.

Figure 14 demonstrates the relationship between the average
transverse flow velocity $\langle\beta_t\rangle$ and
$\sqrt{s_{NN}}$ for different particles in panels (a)--(f), where
the results for different centrality intervals are shown. With
increasing $\sqrt{s_{NN}}$, one can see a significantly increasing
$\langle\beta_t\rangle$ for different hadrons in most cases. At
different energies, $\langle\beta_t\rangle$ for each kind of
particles from central to peripheral collisions decreases
gradually. Observing the values of $\langle\beta_t\rangle$, one
can clearly find that the order of flow velocity is
$\overline{p}(p)>K^{\mp}>\pi^{\mp}$.

In order to represent the chaos degree of the final state particle
information, we show the dependence of pseudo-entropy
$S'_{hadron}$ on $\sqrt{s_{NN}}$ and centrality in Figure 15,
where panels (a)--(f) are orderly for the results of $\pi^-$,
$\pi^+$, $K^-$, $K^+$, $\overline p$, and $p$. One can see that
$S'_{hadron}$ of the six kinds of hadrons in different centrality
intervals increases with the increase of $\sqrt{s_{NN}}$. In
particular, for $\pi^{\mp}$ and $K^{\mp}$ and in central
collisions, $S'_{hadron}$ increases quickly and then slowly with
the increase of $\sqrt{s_{NN}}$, and the boundary energy is 7.7
GeV. From central to peripheral collisions, $S'_{hadron}$
decreases gradually. The chaos degree of particle information at
low energy is less than that at high energy. In addition, the
chaos degree in central collisions is larger than that in
peripheral collisions. Observing the values of $S'_{hadron}$, one
can clearly find that the order of chaos degree is
$\overline{p}(p)>K^{\mp}>\pi^{\mp}$.

In the above discussion, we have used a particular
parameterization for the $p_T$ distribution and obtained the
tendencies of parameters, but many other parameterizations are
available and used in the field. Our conclusions on the behaviors
of main parameters $n$ [or $q=(n+1)/n$], $T_0$, and
$\langle\beta_t\rangle$ with increasing the collision energy,
event centrality, and particle mass are similar to others if one
uses different parameterizations~\cite{54}, though the absolute
sizes may be different. The parameter $a_0$ is a new one
introduced by us recently~\cite{44}, which is hard to compare with
others at present due to the lack of related work, though the
relative sizes of $a_0$ on particle mass are understandable. The
parameterization used in this paper is also suitable for smaller
system studied in our previous work~\cite{44,46}, where the
dependences of parameters on the collision energy, event
centrality (or multiplicity), and particle mass are similar, which
is not very sensitive to the system size.


\begin{figure*}
\begin{center}
\includegraphics[width=14.0cm]{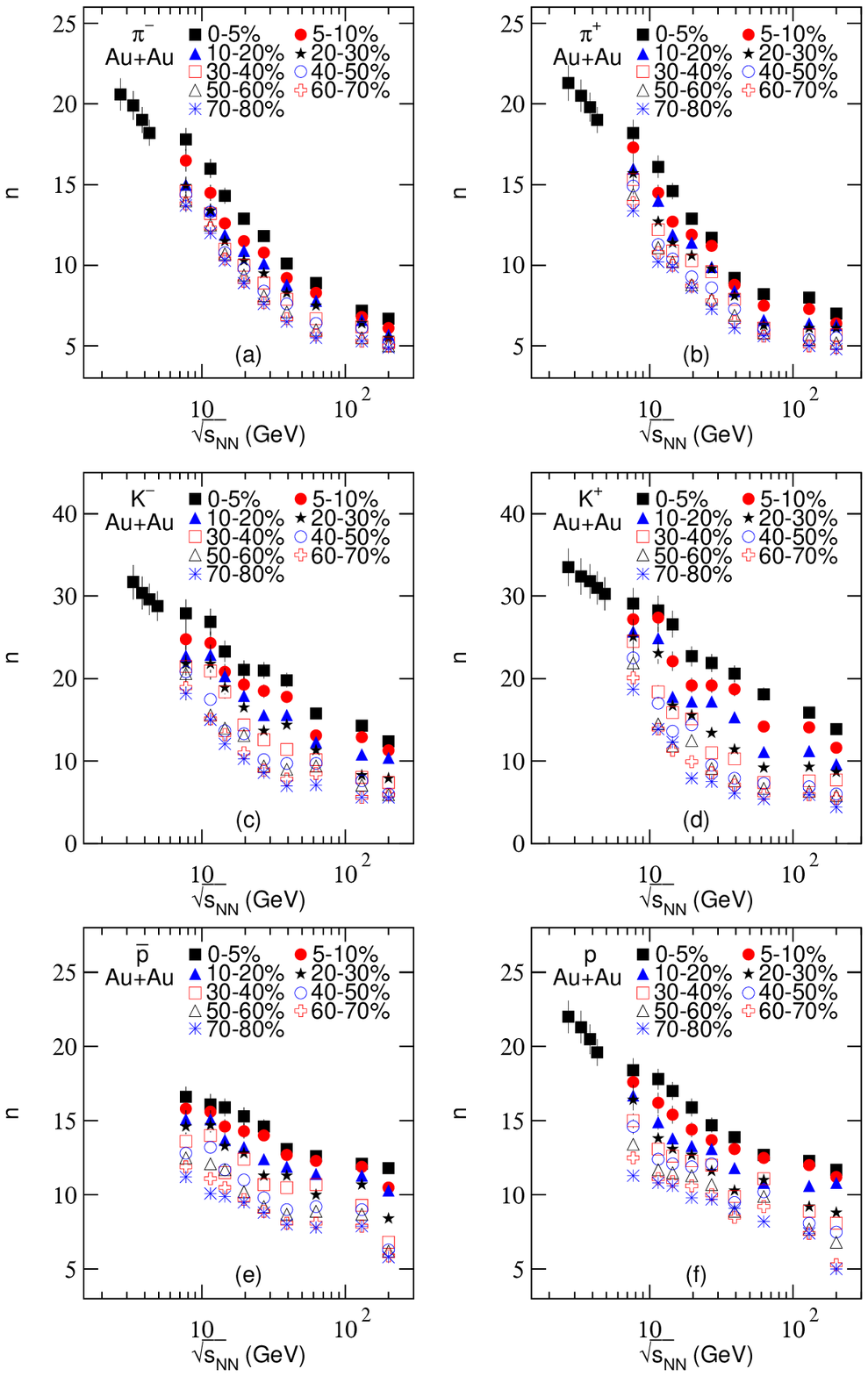}
\end{center}
\vskip-0.4cm \justifying\noindent {Figure 11. Dependence of power
index $n$ on collision energy $\sqrt{s_{NN}}$ in Au-Au collision
with nine main centrality intervals which are marked in the
panels. Figures 11(a)--11(f) correspond to the results from
$\pi^-$, $\pi^+$, $K^-$, $K^+$, $\overline{p}$, and $p$ spectra,
respectively. In particular, at 130 GeV, the centrality intervals
0--6$\%$, 6--11$\%$, 11--18$\%$, 18--26$\%$, 26--34$\%$,
34--45$\%$, and 45--58$\%$ are orderly classified into the closest
main centrality intervals, and the centrality interval 58--85\% is
simultaneously classified into 60--70\% and 70--80\%. The results
for different particles are cited from Tables 1--10.}
\end{figure*}

\begin{figure*}
\begin{center}
\includegraphics[width=14.0cm]{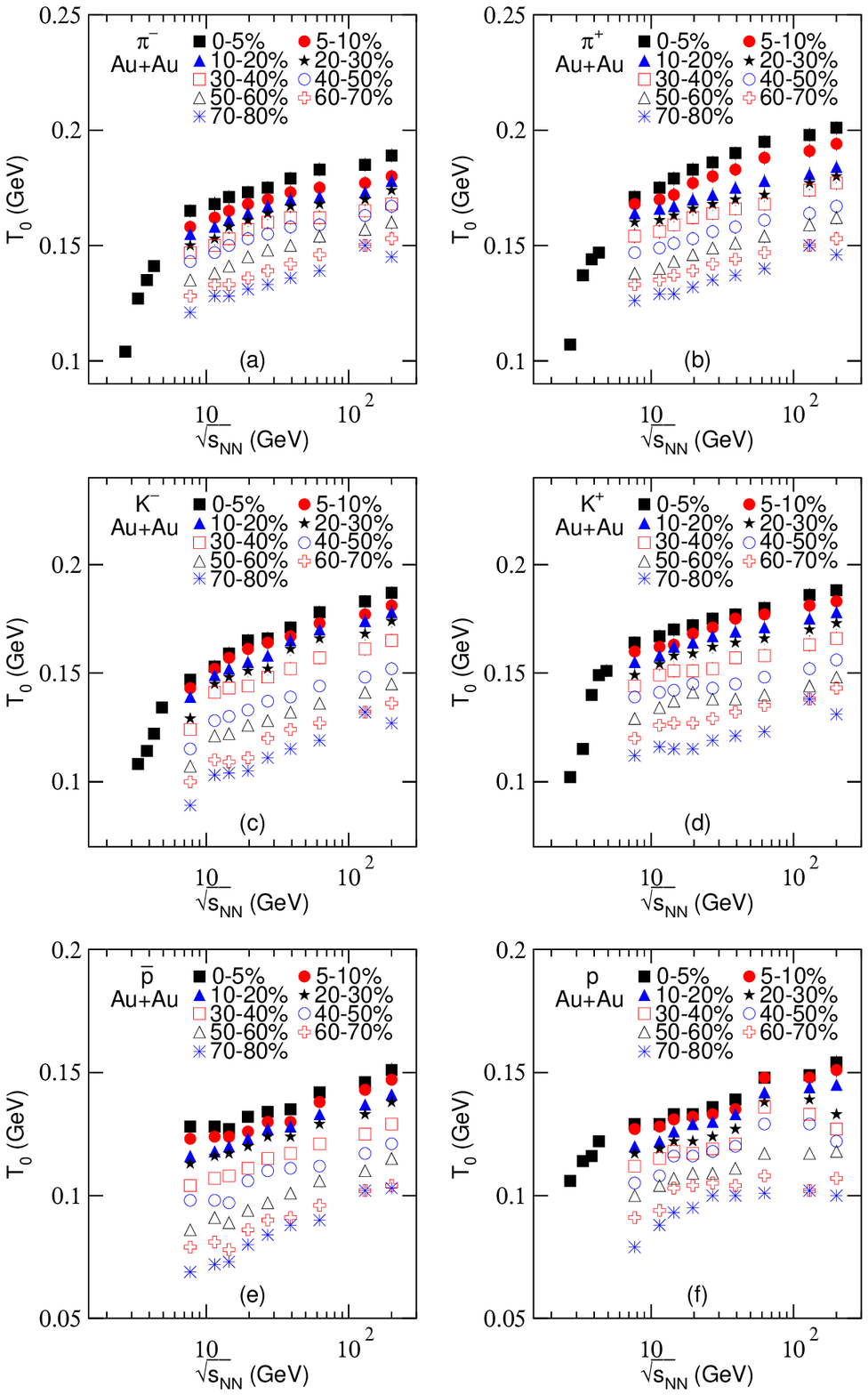}
\end{center}
\justifying\noindent {Figure 12. Same as Figure 11, but showing
the dependence of thermal freeze-out temperature $T_{0}$ on
$\sqrt{s_{NN}}$.}
\end{figure*}

\begin{figure*}
\begin{center}
\includegraphics[width=14.0cm]{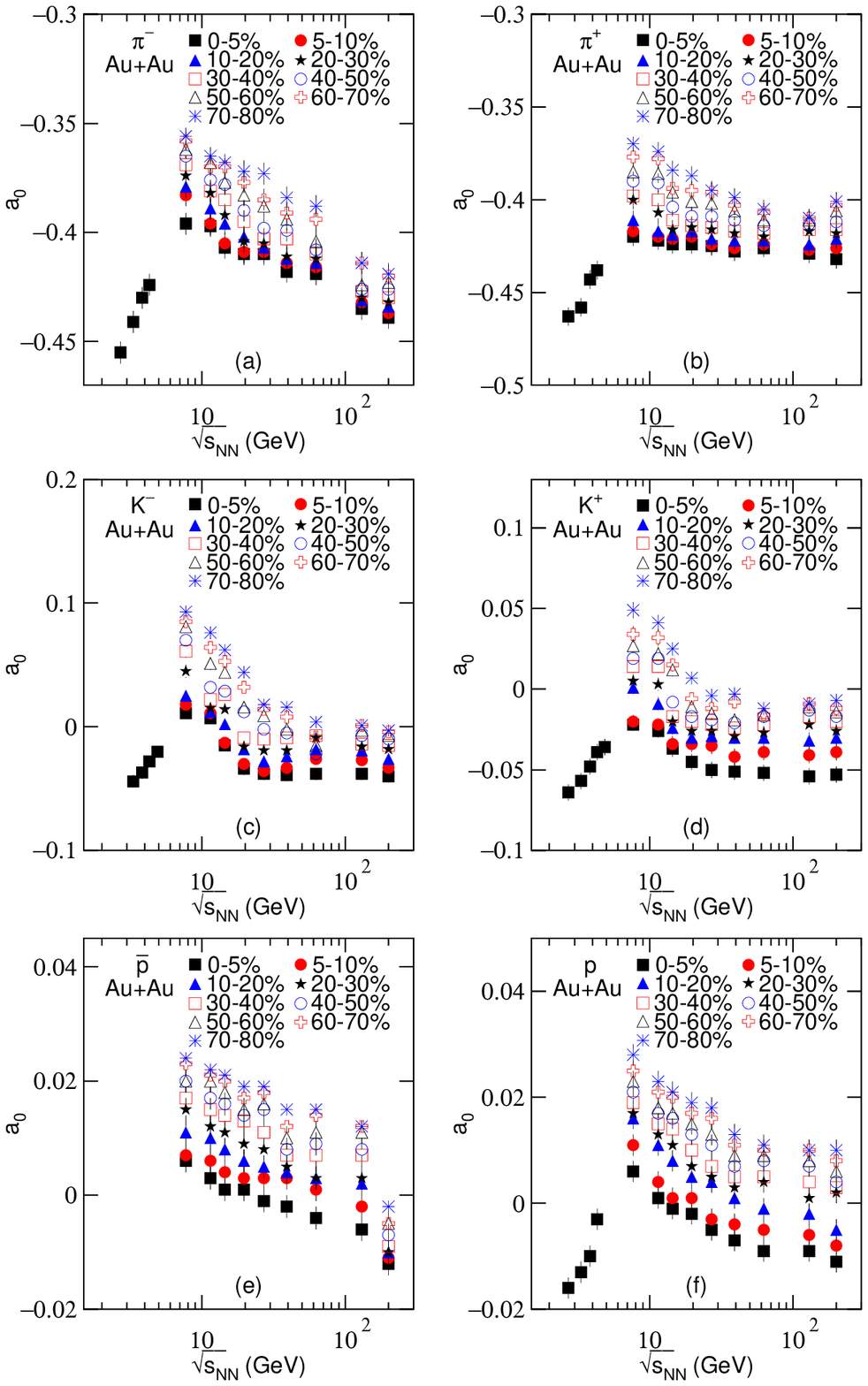}
\end{center}
\justifying\noindent {Figure 13. Same as Figure 11, but showing
the dependence of correction index $a_{0}$ on $\sqrt{s_{NN}}$.}
\end{figure*}

\begin{figure*}
\begin{center}
\includegraphics[width=14.0cm]{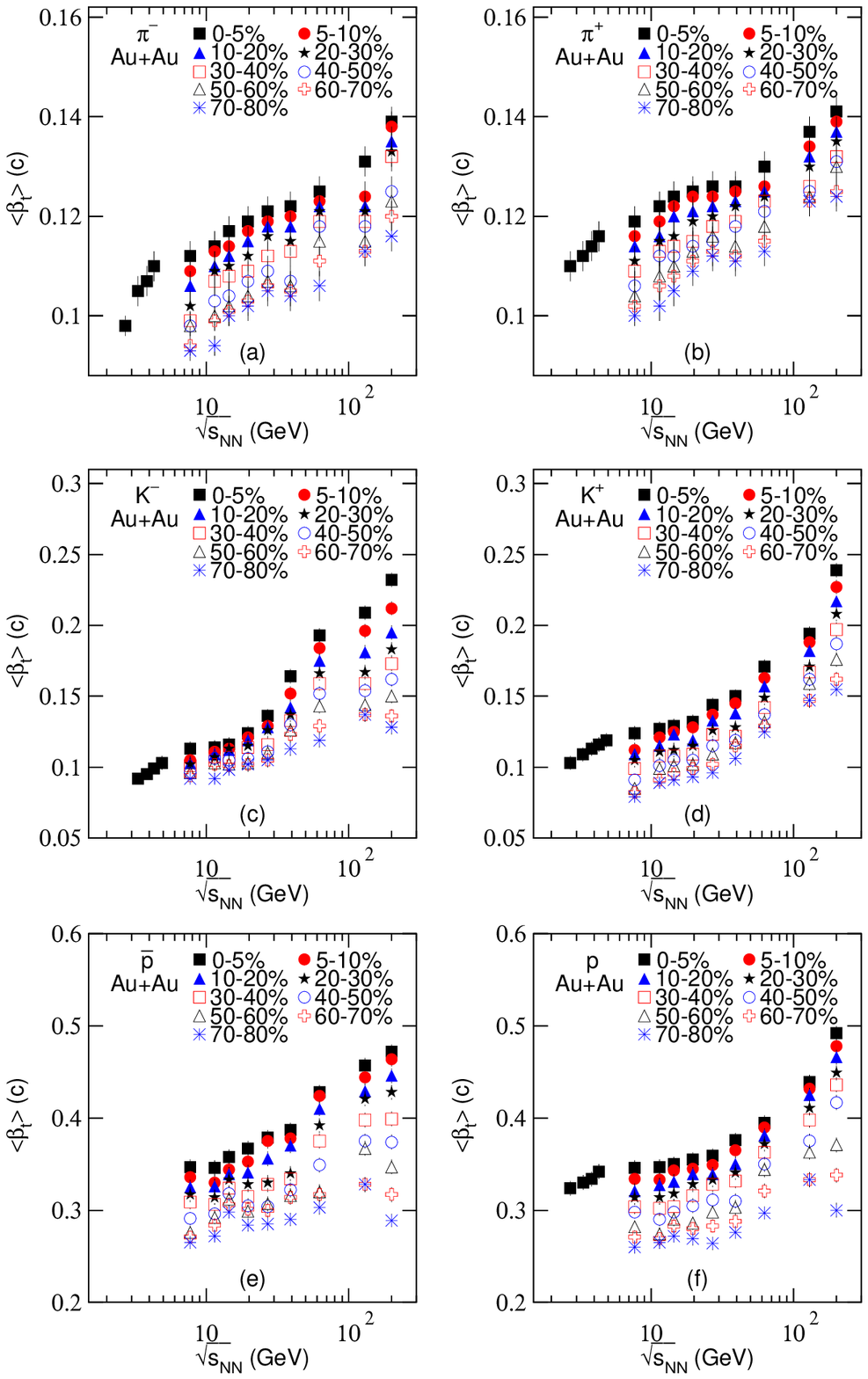}
\end{center}
\justifying\noindent {Figure 14. Same as Figure 11, but showing
the dependence of average transverse flow velocity $\langle
\beta_t \rangle$ on $\sqrt{s_{NN}}$.}
\end{figure*}

\begin{figure*}
\begin{center}
\includegraphics[width=14.0cm]{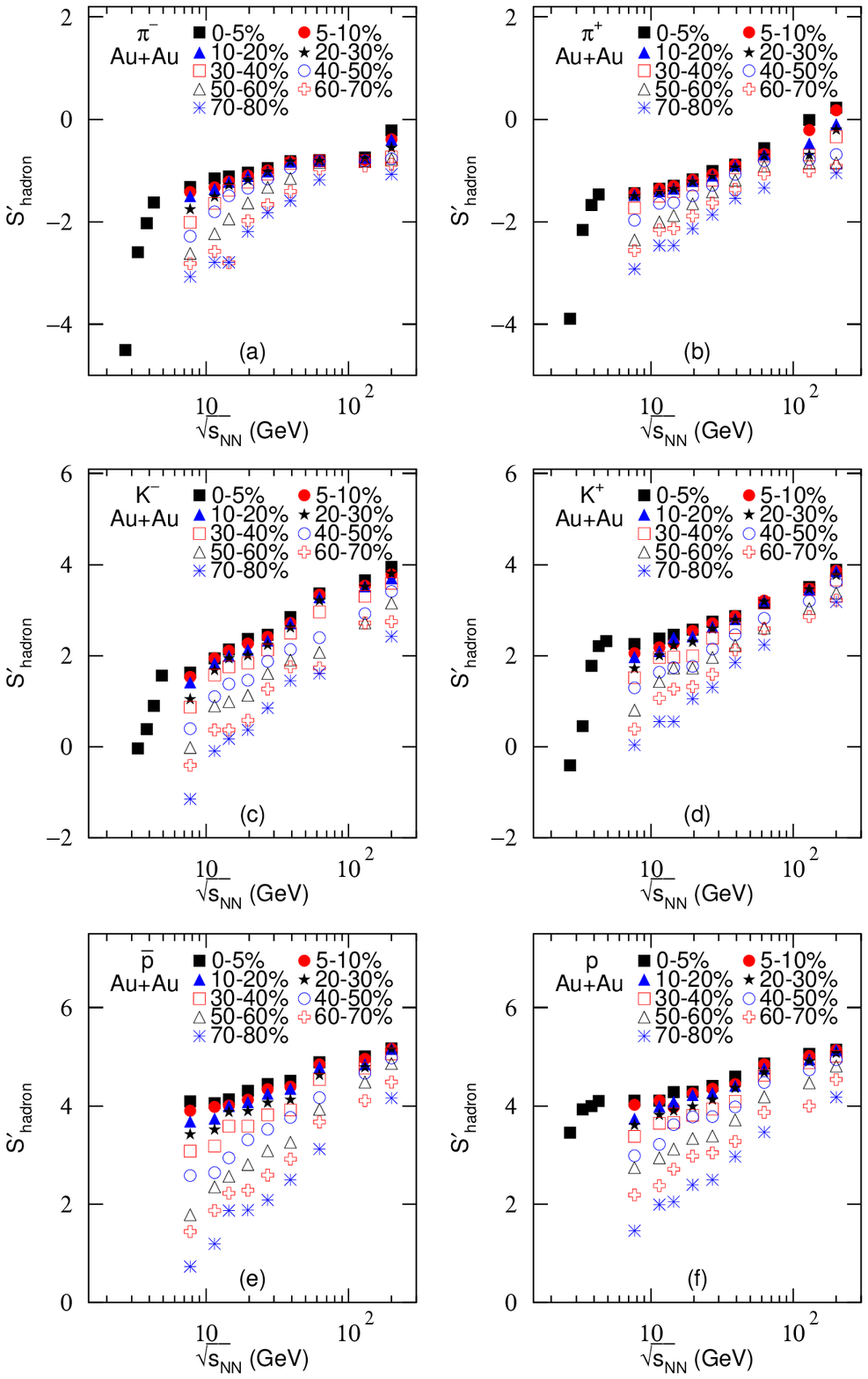}
\end{center}
\justifying\noindent {Figure 15. Same as Figures 11 and 15, but
showing the dependence of pseudo-entropy $S'_{hadron}$ on $C$.}
\end{figure*}

\subsection{Further discussions}

We now discuss this topic further. As we know, high energy
collisions contain abundant information which are related to the
productions of QGP and
particles~\cite{18a,18b,18c,18d,18e,18f,18g,18h}. Of course, after
a very short time, QGP finally decays into particles. The
final-state products are naturally various kinds of particles.
Indeed, understanding the characteristics of particle productions
is particularly necessary for researchers to study the evolution
of collision system and interactions among various
particles~\cite{21a,21b,21c,21d,21e,21f,21g,21h}. By studying the
particle spectra, this paper has extracted some quantities.

From the above comparison, one can see that this paper has used
the transverse mass which contains the average transverse flow
velocity $\langle\beta_t\rangle$ through the Lorentz-like
transformation,
$m'_{T}\rightarrow\langle\gamma_{t}\rangle(m_{T}-p_{T}
\langle\beta_{t}\rangle)$~\cite{62}. Considering the influence of
transverse flow, the effective temperature $T$ in Eq. (1) has been
changed to the thermal freeze-out temperature $T_0$ in Eq. (2). In
this way, $T_0$ and $\langle\beta_t\rangle$ can be directly
extracted~\cite{62}. In this work, $T_0$ and
$\langle\beta_t\rangle$ are quantities at the quark level, but not
the quantities at the particle level. This renders that the values
obtained at the two levels are different.

There is a relation between the power index $n$ and the entropy
index $q$, $n=1/(q-1)$. It shows indeed that a larger $n$
corresponds to a smaller $q$ that is also closer to 1. As we know,
an exact and concrete $q$ that separates an approximate equilibrium
and non-equilibrium states is not available in literature.
Empirically, we may use $q=1.25$, i.e. $n>4$, as the boundary
between the approximate equilibrium and non-equilibrium states.
The present work shows that $n>4$ in most cases. This means that
the system in terms of quarks or partons discussed here is at
the approximate equilibrium or local equilibriums. The concept
of temperature can be approximately used.

At lower energy and in central collisions, the system consisted of
quarks or partons is at larger degree of equilibrium. This is
understood in view of the longer reaction time in lower collision
energy and more participant partons in central collisions. These
two factors can affect the system to be at more equilibrium due to
more cascade collisions happening. At the present energy at the
LHC, the system is approximately at the equilibrium. It is
possible that at very high energy the system will be at
non-equilibrium due to very short reaction time. In most cases,
the system in peripheral collisions is approximately at the
equilibrium. It is not preclusive that in very peripheral
collisions the system is possible at non-equilibrium due to very
less cascade collisions in very limited collision volume, where
the multiplicity is also low.

There are possible scenarios for single, double, or multiple
kinetic freeze-out in literature. The single (two or multiple)
kinetic freeze-out means that a set (two or multiple sets) of
parameters can fit the spectra. There is not a sold conclusion for
a definite scenario. The present work shows that some parameters
from different particle spectra are different. This reflects the
scenario for multiple kinetic freeze-out in high-energy
collisions. According to the present work, we may say that
different particles are produced in the system at different times.
The decay of the system is not all at once, but successive.

The value of $T_0$ reflects the excitation degree of the system at
the parton level, which is somehow larger than that at the
particle level, due to the former happening earlier. The higher
the value of $T_0$ is, the higher the excitation degree of the
system is, and the larger the deposited energy is. The value of
$\langle\beta_t\rangle$ reflects impact and squeeze between the
projectile and target at parton level. In high energy collisions,
a lot of energy are deposited in the process of parton
interactions, though some partons go through the system as
spectators.

Some soft excitation and hard scattering processes among partons
have happened. The contribution of soft component from resonance
in $\pi^{\mp}$ spectra is significantly larger than those in
$K^{\mp}$ and $\overline p(p)$ spectra. This renders that the
spectra of $\pi^{\mp}$ have a significant increase in low-$p_T$
region, comparing with the spectra of other particles. This change
in low-$p_T$ region is mainly described by the value of $a_0$ in
this work. A small $a_0$ describes a significant increase and a
large $a_0$ describes a decrease. Naturally, this change also
affects the parameters $n$, $T_0$, and $\langle\beta_t\rangle$.

Indeed, for $\pi^{\mp}$ spectra in low-$p_T$ region, the
production of resonances cannot be ignored. More contributions
from the resonances in $\pi^{\mp}$ productions render that the
shape of the spectra of the considered different kinds of
particles are different, in particular in low-$p_T$ region. This
also renders the different sets of parameters for different
particles. In addition, multiple kinetic freeze-out also results
in multiple sets of parameters. The earlier the particles are
emitted, the wider the spectrum is, and the higher the temperature
is.

At given collision energy, the nuclear environments (the sizes of
participant or spectator regions) in central and peripheral Au-Au
collisions have some influences on the transverse momentum
spectra. We attribute this phenomenon to the hadronic cascade
collisions (or multiple scattering) in hot and dense participant
zone. The larger number of participants in central collisions
means more cascade collisions and, hence, the larger transverse
momentum and temperature. In the considered transverse momentum
region, the changing laws of transverse momentum spectra for given
particles are independent of isospin. Compared with strong
interactions, contribution of electromagnetic interactions is very
small and negligible.

The purpose of the RHIC-BES program is to search for the critical
point in the QCD phase diagram. In the RHIC-BES and the related
AGS energy regions, from the energy dependences of related
parameters in the available central collisions, we found that
$T_0$ and $S'_{hadron}$ increase quickly and then slowly at around
7.7 GeV with the increase of collision energy. This special energy
is the knee point of the equation of state (EoS) of the hot and
dense matter, which is related to the critical energy, if not
equal. Meanwhile, we found that $a_0$ increases and then decreases
at around 7.7 GeV in the available central collisions. Generally,
different parameters may reveal different knee points. The most
probable knee point from the energy dependences of various
parameters can be regarded as the critical point.

In the available central collisions, both the boundary energies of
the rapid-slow increase in the energy dependence of $T_0$ and
$S'_{hadron}$ and of the increase-decrease change in the energy
dependence of $a_0$ are 7.7 GeV. We argue that this energy is
possibly the critical energy of deconfinement phase transition
from hadronic matter to QGP. Because of the complexity of the collision
process, the conclusion about the critical energy in this work is not
solid. A shift of 2--3 GeV is possiable. To make a solid conclusion,
one needs a comprehensive analysis of various parameters. At least,
we may state that the interaction mechanisms or strengths at the
energy below and above about 7.7 GeV are different.

\section{Summary and Conclusions}

In this paper, the TP-like function has been used to analyze the
transverse momentum spectra of the charged particles [$\pi^{\mp}$,
$K^{\mp}$, and $\overline{p}(p)$] produced in Au-Au collisions at
the RHIC-BES energies and the related AGS energies. The
contribution of each constituent quark to the transverse momentum
of charged particle is assumed to satisfy the TP-like function.
Each constituent quark is also regarded as an energy resource.
Therefore, the transverse momentum spectrum of mesons (baryons) is
the convolution of two (three) TP-like functions. According to the
fitting results, we have obtained four parameters: the power index
$n$, the thermal freeze-out temperature $T_{0}$, the correction
index $a_{0}$, and the average transverse flow velocity
$\langle\beta_{t}\rangle$. The dependences of these parameters on
collision energy and centrality have been also investigated.

It is obvious that with the increase of collision energy, $n$
decreases, which indicates that the collision system is farther
away from the equilibrium, though it is still approximately at the
equilibrium, at higher energy. In other words, at lower energy,
the longer the collision evolves, the closer to the equilibrium
the system is. More importantly, with the increase of collision
energy, $T_0$ extracted from meson spectra in central Au-Au
collisions increases quickly in the range of less than 7.7 GeV and
slowly in the range of greater than 7.7 GeV. Meanwhile,
$\langle\beta_{t}\rangle$ increases with the increase of collision
energy. From central to peripheral collisions, all of $n$,
$T_{0}$, and $\langle\beta_{t}\rangle$ decrease, and $a_0$
increases. Even in peripheral collisions, the system is
approximately at the equilibrium.

We have introduced the concept and definition of a new quantity,
the pseudo-entropy. The pseudo-entropy is carried out a more
in-depth study on the transverse momentum spectra of charged
particles. With the increase of collision energy, the
pseudo-entropy extracted from the meson spectra in central Au-Au
collisions increases quickly in the range of less than 7.7 GeV and
slowly in the range of greater than 7.7 GeV. This behavior is
similar to that for $T_0$. We argue that 7.7 GeV is a special
energy at which the phase transition of deconfinement from
hadronic matter to QGP can possibly happen. At least, the
interaction mechanism in the system or the production mechanism of
charged particles at the energy below and above about 7.7 GeV
are different.
\\
\\
{\bf Data Availability}

The data used to support the findings of this study are included
within the article and are cited at relevant places within the
text as references.
\\
\\
{\bf Ethical Approval}

The authors declare that they are in compliance with ethical
standards regarding the content of this paper.
\\
\\
{\bf Disclosure}

The funding agencies have no role in the design of the study; in
the collection, analysis, or interpretation of the data; in the
writing of the manuscript; or in the decision to publish the
results.
\\
\\
{\bf Conflicts of Interest}

The authors declare that there are no conflicts of interest
regarding the publication of this paper.
\\
\\
{\bf Acknowledgments}

The work of X.H.Z. and F.H.L. was supported by the National
Natural Science Foundation of China under Grant Nos. 12147215,
12047571, 11575103, and 11947418, the Scientific and Technological
Innovation Programs of Higher Education Institutions in Shanxi
(STIP) under Grant No. 201802017, the Shanxi Provincial Natural
Science Foundation under Grant No. 201901D111043, and the Fund for
Shanxi ``1331 Project" Key Subjects Construction. The work of
X.H.Z. was also supported by the Innovative Foundation for
Graduate Education in Shanxi University. The work of Y.Q.G. was
supported by the Scientific and Technological Innovation Programs
of Higher Education Institutions in Shanxi (STIP) under Grant No.
2019L0629. The work of K.K.O. was supported by the Ministry of
Innovative Development of the Republic of Uzbekistan within the
fundamental project No. F3-20200929146 on analysis of open data on
heavy-ion collisions at RHIC and LHC.
\\
\\

\end{document}